\documentclass[english,jcp,preprint,endfloats,groundaddress]{revtex4-1}

\usepackage[T1]{fontenc}
\usepackage{graphicx}
\usepackage{amsfonts}
\usepackage{amssymb}
\usepackage{longtable}
\usepackage{multirow}
\usepackage{supertabular}
\usepackage{epsfig}

\makeatletter

\begin{document}

\title{Density functional theory with fractional orbital occupations}

\author{Jeng-Da Chai}
\email{Electronic mail: jdchai@phys.ntu.edu.tw.}
\address{Department of Physics, Center for Theoretical Sciences, and Center for Quantum Science and Engineering, National Taiwan University, Taipei 10617, Taiwan}

\date{\today{}}

\begin{abstract}

In contrast to the original Kohn-Sham (KS) formalism, we propose a density functional theory (DFT) with fractional orbital occupations for the study of ground states of many-electron systems, 
wherein strong static correlation is shown to be described. Even at the simplest level represented by the local density approximation (LDA), 
our resulting DFT-LDA is shown to improve upon KS-LDA for multi-reference systems, such as dissociation of H$_2$ and N$_2$, and twisted ethylene, while performing similarly to KS-LDA for 
single-reference systems, such as reaction energies and equilibrium geometries. Because of its computational efficiency (similar to KS-LDA), this DFT-LDA is applied to the study of the 
singlet-triplet energy gaps (ST gaps) of acenes, which are "challenging problems" for conventional electronic structure methods due to the presence of strong static correlation effects. 
Our calculated ST gaps are in good agreement with the existing experimental and high-level {\it ab initio} data. The ST gaps are shown to decrease monotonically with the increase of chain length, 
and become vanishingly small (within 0.1 kcal/mol) in the limit of an infinitely large polyacene. In addition, based on our calculated active orbital occupation numbers, the ground states for 
large acenes are shown to be polyradical singlets. 

\end{abstract}

%\pacs{}
\maketitle
\newpage

\section{Introduction}

As the problem of solving the $N$-electron Schr\"odinger equation quickly becomes intractable as the size of a system increases, the development of efficient and accurate electronic structure 
methods for large systems, continues being the subject of intense current interest. Over the past two decades, Kohn-Sham density functional theory (KS-DFT) \cite{HK,KS} has become one of the 
most popular theoretical approaches for calculations of electronic properties of large ground-state systems (up to a few thousand electrons), due to its favorable balance between cost and 
performance \cite{Parr,Kohanoff,Jensen,DFTReview,Perdew09}. Recently, its time-dependent extension, time-dependent density functional theory (TDDFT), has also been actively developed for treating electron 
dynamics and excited states of large systems with considerable success \cite{Casida,Gross}.  

Although the exact exchange-correlation (XC) functional $E_{xc}[\rho]$ in KS-DFT has not been known, functionals based on the standard density functional approximations (DFAs), such as the 
local density approximation (LDA) and generalized gradient approximations (GGAs), can accurately describe short-range XC effects (due to the accurate treatment of on-top hole density), and are 
computationally favorable for large systems \cite{Parr,Kohanoff,Jensen,DFTReview,Perdew09}. Although KS-DFAs have been successful in many applications, due to the lack of accurate treatment of nonlocality of XC hole, 
KS-DFAs can exhibit the following three types of qualitative failures: (i) self-interaction error (SIE), (ii) noncovalent interaction error (NCIE), and (iii) static correlation error (SCE). 
In situations where these failures occur, KS-DFAs can produce erroneous results \cite{SciYang}. Therefore, resolving the qualitative failures of KS-DFAs at a reasonable computational cost seems to be 
the first step toward finding an efficient and accurate electronic structure method for large systems. 

The SIEs of KS-DFAs lead to drastic failures for problems such as barrier heights of chemical reactions, band gaps of solids, and dissociation of symmetric radical cations \cite{SIE}. In TDDFT, 
SIE causes failures for problems such as Rydberg excitations in molecules and long-range charge-transfer excitations between two well-separated molecules \cite{Dreuw}. The SIEs of KS-DFAs can be 
greatly reduced by hybrid DFT methods \cite{hybrid}, incorporating some of the exact Hartree-Fock (HF) exchange into the KS-DFAs. Over the past twenty years, several hybrid functionals, such as 
global hybrid functionals \cite{B3LYP,M06-2X} and long-range corrected (LC) hybrid functionals \cite{LC-DFT,LCHirao,CAM-B3LYP,LC-wPBE,wB97X}, have been developed to improve the accuracy of 
$E_{xc}[\rho]$, extending the applicability of KS-DFT to a wide range of systems.  

The proper treatment of noncovalent interactions requires the accurate description of dynamical correlation effects at medium and long ranges, which cannot be properly captured by KS-DFAs \cite{Dobson2}. 
In particular, for dispersion-dominated (van der Waals (vdW)) interactions, KS-LDA tends to overestimate the binding energies, while KS-GGAs tend to give insufficient binding or even unbound 
results. The NCIEs of KS-DFAs can be efficiently reduced by the DFT-D (KS-DFT with empirical dispersion corrections) schemes \cite{DFT-D1,DFT-D2,wB97X-D,DFT-D3}, which have shown 
generally satisfactory performance on a large set of noncovalent systems \cite{Sherrill,DFT-D3appl}. The dispersion corrections can also be computed less empirically by the exchange-hole dipole 
moment (XDM) method \cite{BDFTD} or by the local response dispersion (LRD) method \cite{LRD}. Alternatively, a fully nonlocal density functional for vdW interactions (vdW-DF) \cite{vdW} can also 
be adopted to reduce the NCIEs of KS-DFAs. Currently, perhaps the most successful approach to taking into account nonlocal dynamical correlation effects is provided by the double-hybrid (DH) 
methods \cite{B2PLYP,XYG3,wB97X-2}, which mix some of the HF exchange and some of the nonlocal orbital correlation energy from the second-order M\o ller-Plesset perturbation (MP2) 
theory \cite{MP2} into the KS-DFAs. DH functionals have shown an overall satisfactory accuracy for thermochemistry, kinetics, and noncovalent interactions. In addition, the sharp increase in HF exchange 
in typical DH functionals has also greatly reduced the SIEs relative to KS-DFAs and conventional hybrid functionals. 

Systems with strong static (nondynamical) correlation effects, such as bond-breaking reactions, diradicals, conjugated polymers, magnetic materials, and transition-metal compounds, belong to the class 
of strongly correlated (SC) systems (multi-reference systems). Such a system is usually characterized by a small (or vanishing) energy gap between the highest occupied molecular orbital (HOMO) and 
the lowest unoccupied molecular orbital (LUMO), the HOMO-LUMO gap. Despite their computational efficiency, the accurate treatment of SC systems poses a great challenge to KS-DFAs and 
hybrid DFT methods \cite{SciYang,SCE}. The DH methods may also be inadequate for SC systems, as the second-order perturbation energy components diverge to minus infinity for systems with vanishing 
HOMO-LUMO gaps. Within the framework of KS-DFT, fully nonlocal XC functionals, such as those based on random phase approximation (RPA), are believed to be essential for the accurate treatment 
of SC systems \cite{DFTReview,Perdew09}. However, compared with KS-DFAs, hybrid functionals, and DH functionals, RPA-type functionals are computationally very demanding for large systems, and 
their applications to SC systems are too scarce to make a firm judgment on their accuracy. Recently, the SIEs of RPA-type functionals have been shown to be severe, even for a simple one-electron 
system such as H$_{2}^{+}$ \cite{RPAf}. 

Aiming to reduce the SCEs of KS-DFAs without extra computational cost, we propose a DFT with fractional orbital occupations, rather than a fully nonlocal XC functional in KS-DFT. 
The rest of this paper is organized as follows. In section II, we briefly describe the rationale for fractional orbital occupations. In section III, we describe the formulation of this DFT and explain how 
strong static correlation is described by this DFT. At the simple LDA level, the performance of our resulting DFT-LDA is compared with KS-LDA for several single- and multi-reference systems in section IV. 
Based on physical arguments and numerical investigations, the optimal parameter for this DFT-LDA is defined in section V. Our conclusions are given in section VI.

\section{Rationale for Fractional Orbital Occupations}

For the exact DFT, the exact ground-state energy can be obtained, only when the exact ground-state density is available to insert into the exact ground-state energy functional. Therefore, the development of a 
generally accurate DFT method should involve not only an accurate approximation for the ground-state energy functional, but also an appropriate representation (possibly in terms of orbitals and occupation 
numbers) of the ground-state density. However, much less attention has been paid to the latter (representation of ground-state density) than to the former (ground-state energy functional). 
Due to the search over a restricted domain of densities, the exact ground-state density of interest may not be obtained within the framework of KS-DFT, in which case even the exact KS-DFT will 
fail (i.e.\ the exact XC functional may not be differentiable at the exact ground-state density) \cite{Perdew09}. Noticeably, some of these situations occur for systems with vanishingly small HOMO-LUMO gaps (SC systems), 
indicating the close relationship between strong static correlation effects and representations of the ground-state density. Therefore, to accurately describe SC systems, it seems intuitive to focus on devising an appropriate 
representation for the exact ground-state density and a DFT associated with such a representation. 

A ground-state density $\rho({\bf r})$ is said to be interacting $v$-representable, if it can be obtained from a ground-state wave function of an interacting $N$-electron Hamiltonian for some external 
potential $v({\bf r})$. The exact $\rho({\bf r})$ can be obtained by the full configuration interaction (FCI) method at the complete basis set limit (i.e.\ interacting $v$-representable) \cite{H2_NOON}, and 
can be compactly expressed in terms of the natural orbitals (NOs) $\{\chi_{i}({\bf r})\}$ and natural orbital occupation numbers (NOONs) $\{n_{i}\}$ \cite{NO}:
\begin{equation}
\rho_{\text{FCI}}({\bf r}) = \sum_{i=1}^{\infty} n_{i} |\chi_{i}({\bf r}) |^{2},
\label{eq:rhoci}
\end{equation}
where $\{n_{i}\}$, obeying the following two conditions,  
\begin{equation}
\sum_{i=1}^{\infty} n_{i} = N,\  \ 0\le n_{i} \le 1,
\label{eq:rhoci2}
\end{equation}
are related to the variationally determined coefficients of the FCI expansion. As shown in Eq.\ (\ref{eq:rhoci}), the exact $\rho({\bf r})$ can be represented by orbitals and occupation numbers, highlighting the importance 
of ensemble-representable densities. 

By contrast, in KS-DFT \cite{KS}, $\rho({\bf r})$ is {\it assumed} to be noninteracting pure-state (NI-PS) $v_s$-representable, as it belongs to a one-determinantal ground-state wave function of 
a noninteracting $N$-electron Hamiltonian (KS Hamiltonian) for some local potential $v_s({\bf r})$ (KS potential) \cite{v-rep,Levy,Lieb}. Correspondingly, the KS orbital occupation numbers should be 
either 0 or 1. As the Aufbau principle (i.e.\ filling the KS orbitals in order of increasing energy) should be obeyed, $\rho({\bf r})$ can be expressed as the sum of the densities of the lowest $N$ 
occupied KS orbitals $\{\phi_{i}({\bf r})\}$ \cite{Parr}: 
\begin{equation}
\rho_{\text{KS-DFT}}({\bf r}) = \sum_{i=1}^{N} |\phi_{i}({\bf r}) |^{2}.
\label{eq:rhoks}
\end{equation}
As ground-state densities of most nondegenerate atomic and molecular systems (e.g.\ closed-shell systems with sizable HOMO-LUMO gaps) are likely to be NI-PS $v_s$-representable, the commonly 
used XC functionals in KS-DFT are reliably accurate for these systems (assuming that the SIEs and NCIEs of these functionals are not severe). 
However, if a one-determinantal ground-state wave function is insufficient to represent $\rho({\bf r})$, these XC functionals may not be reliably accurate, as they 
are all developed based on general properties of systems where the KS wave function is a one-determinantal wave function \cite{Perdew09}. 

As shown by several researchers \cite{Levy,Lieb,Morrison,Katriel,Baerends}, there are some reasonable ground-state densities that are {\it not} NI-PS $v_s$-representable. Clearly, such densities cannot 
be obtained within the framework of KS-DFT. To remedy this situation, KS-DFT has been extended to ensemble DFT (E-DFT), wherein $\rho({\bf r})$ is {\it assumed} to be noninteracting ensemble 
(NI-E) $v_s$-representable, as it belongs to an ensemble of pure determinantal states of the noninteracting KS system \cite{EDFT,EDFT2}. Correspondingly, $\rho({\bf r})$ can be expressed as 
\begin{equation}
\rho_{\text{E-DFT}}({\bf r}) =  \sum_{i=1}^{\infty} g_{i} |\phi_{i}({\bf r}) |^{2},
\label{eq:rhoks2}
\end{equation}
where the occupation numbers $\{g_{i}\}$, obeying the following two conditions, 
\begin{equation}
\sum_{i=1}^{\infty} g_{i} = N,\  \ 0\le g_{i} \le 1,
\label{eq:rhoks2a}
\end{equation}
are given by   
\begin{equation}
g_{i} = \left\{ \begin{array}{rcl}
1, & \mbox{for} & \epsilon_{i} < \epsilon_{F} \\ 
x_{i}, & \mbox{for} & \epsilon_{i} =  \epsilon_{F} \\
0, & \mbox{for} & \epsilon_{i} > \epsilon_{F}.
\end{array}\right.
\label{eq:rhoks3}
\end{equation}
Here $\epsilon_{i}$ is the orbital energy of the $i^{\text{th}}$ KS orbital $\phi_{i}({\bf r})$, $\epsilon_{F}$ is the Fermi energy, and $x_{i}$ is a fractional number (between 0 and 1). 
As can be seen, fractional orbital occupations can occur only for the orbitals at the Fermi level. 

In 1998, the close relationship between strong static correlation effects and non-NI-PS $v_s$-representable densities was observed by Baerends and co-workers \cite{Baerends}, who studied 
the $^1\Sigma_{g}^{+}$ ground state of the C$_2$ molecule (a system where the ground-state wave function is nondegenerate but has a strong multi-reference character), and investigated the 
possibility that the ground-state density $\rho({\bf r})$ may be NI-E $v_s$-representable (as assumed in E-DFT), rather than NI-PS $v_s$-representable (as assumed in KS-DFT). In their study, 
$\rho({\bf r})$ was first obtained from the highly accurate {\it ab initio} CI wave function method, wherein $\rho({\bf r})$ can be represented by the NOs $\{\chi_{i}({\bf r})\}$ and 
NOONs $\{n_{i}\}$ in Eq.\ (\ref{eq:rhoci}). An iterative method for the construction of the KS orbitals and the KS potential from the CI density was then adopted \cite{LB94}, combined with the 
constraint of integer occupations of the KS orbitals. The $\rho({\bf r})$ of C$_2$ was found to be NI-PS $v_s$-representable at a bond distance shorter than the equilibrium distance, 
while being non-NI-PS $v_s$-representable at the longer bond distances, leading to non-Aufbau solutions with unoccupied KS orbitals having energies lower than those of the highest occupied KS 
orbitals (i.e.\ holes below the Fermi level). On the other hand, when the $\rho({\bf r})$ of C$_2$ was represented by the ensemble solution during the above iterative procedure, no holes below 
the Fermi level were found, and the corresponding KS orbitals were close to the NOs. Based on their results, Baerends and co-workers \cite{Baerends} argued that the KS orbitals 
(generated from Aufbau solutions) for a NI-PS $v_s$-representable ground-state density are comparable to the NOs , while the KS orbitals (generated from non-Aufbau solutions) for 
a non-NI-PS $v_s$-representable ground-state density can be distinctly different from the NOs in order to incorporate the effect of the configuration mixing on the ground-state density. 
They concluded that the ground-state density of a system with strong static correlation effects is likely to be non-NI-PS $v_s$-representable, in which case an ensemble representation 
(via fractional orbital occupations) of the ground-state density is crucial. Arguments in support of this are also available elsewhere \cite{Morrison,SDFTYang}. 

The idea of simulating strong static correlation effects by fractional occupation numbers (FONs) in DFT has spurred the development of the DFT-FON method \cite{DFT-FON,DFT-FON2,DFT-FON3,
DFT-FON4,SDFTYang}, the spin-restricted ensemble-referenced KS (REKS) method \cite{REKS,REKS2}, and the fractional-spin DFT (FS-DFT) method \cite{SciYang,SCE}, with great success for some SC systems. 
The practical implementation of E-DFT and related methods has, however, been impeded by a few factors as follows. First, a double-counting of correlation effects may occur, when the conventional 
approximate XC functional is evaluated with an ensemble density in Eq.\ (\ref{eq:rhoks2}), rather than a NI-PS $v_s$-representable density in Eq.\ (\ref{eq:rhoks}). 
By taking into account the double-counting effects, the XC functional in E-DFT may need to be re-derived. Second, the computational cost of E-DFT is more expensive than that of KS-DFT, 
which makes E-DFT less practical for the study of large ground-state systems. Third, the computational cost of analytical nuclear gradients (if available) for E-DFT is more expensive than that for KS-DFT, 
which makes it a formidable computational task to perform geometry optimization of large molecules for E-DFT. 

In view of the above difficulties, we focus on the representation of the ground-state density from the exact theory in Eq.\ (\ref{eq:rhoci}). Although the exact orbital occupation numbers for 
interacting electrons are intractable for large systems due to the exponential complexity, the approximate ones can, however, be properly simulated. 
Based on the statistical properties of strongly correlated eigenstates, Flambaum {\it et al.} argued that the distribution of occupation numbers (the microcanonical averaging of the occupation numbers) for 
a finite number of interacting Fermi particles (with two-body interaction) practically does not depend on a particular many-body system and has a universal form that can be approximately described 
by the Fermi-Dirac distribution with renormalized parameters (i.e.\ orbital energies, chemical potential, and temperature) \cite{Flambaum1,Flambaum2,Flambaum3}. 
Statistical effects of the interaction have been shown to be absorbed by introduction of the effective temperature. 

In view of the close relationship between strong static correlation effects and representations of the ground-state density as well as the close relationship between the distribution of occupation numbers 
for interacting electrons (in the sense of statistical average) and the Fermi-Dirac distribution for noninteracting electrons, in this work, the ground-state density $\rho({\bf r})$ of a system of $N$ interacting 
electrons (at zero temperature) in the presence of an external potential $v_{ext}({\bf r})$, is {\it assumed} to be noninteracting thermal ensemble (NI-TE) $v_s$-representable, as it is represented by the 
thermal equilibrium density of an auxiliary system of $N$ noninteracting electrons at a fictitious temperature $\theta \equiv k_{B} T_{el}$ (where $k_{B}$ is the Boltzmann constant, 
$T_{el}$ is the temperature measured in absolute temperature, and $\theta$ is the temperature measured in energy units) in the presence of a local potential $v_s({\bf r})$. 
Correspondingly, $\rho({\bf r})$ can be expressed as 
\begin{equation}
\rho({\bf r}) =  \sum_{i=1}^{\infty} f_{i} |\psi_{i}({\bf r}) |^{2},
\label{eq:rhomks2}
\end{equation}
where the occupation number $f_{i}$ is the Fermi-Dirac function 
\begin{equation}
f_{i} = \{1+\text{exp}[( \epsilon_{i} - \mu)/ \theta] \}^{-1},
\label{eq:mdft5}
\end{equation}
which obeys the following two conditions, 
\begin{equation}
\sum_{i=1}^{\infty} f_{i} = N,\  \ 0\le f_{i} \le 1,
\label{eq:rhomks2a}
\end{equation}
$\epsilon_{i}$ is the orbital energy of the $i^{\text{th}}$ orbital $\psi_{i}({\bf r})$, and $\mu$ is the chemical potential chosen to conserve the number of electrons $N$. 

In the following section, we demonstrate how a DFT associated with the NI-TE $v_s$-representable $\rho({\bf r})$ in Eq.\ (\ref{eq:rhomks2}), can be formulated. In other words, for a given fictitious temperature $\theta$, 
the remaining "renormalized parameters" ($\{\epsilon_{i}\}$ and $\mu$) and the $\{\psi_{i}({\bf r})\}$, can be self-consistently determined to represent the $\rho({\bf r})$ in Eq.\ (\ref{eq:rhomks2}), which then determines 
the ground-state energy of the system. Strong static correlation is shown to be described by a term related to the $\theta$ and $\{f_{i}\}$ in this DFT. 

To avoid any possible confusion with KS-DFT, E-DFT, and finite-temperature DFT (FT-DFT) \cite{Mermin}, we refer to this DFT as thermally-assisted-occupation DFT (TAO-DFT). 
We wish to develop TAO-DFT with the following characteristics. 
\begin{itemize}
\item It is developed for ground-state systems at zero temperature.   
\item It represents the ground-state density with orbitals and occupation numbers.
\item It may be used together with existing XC functionals in KS-DFT. 
\item It reduces to KS-DFT in the absence of strong static correlation effects. 
\item It treats single- and multi-reference systems in a more balanced way than KS-DFT. 
\item It has similar computational cost as KS-DFT (e.g.\ energy, analytical nuclear gradients). 
\end{itemize}

\section{TAO-DFT}

\subsection{Self-Consistent Equations}
 
Consider a system of $N$ interacting electrons moving in an external potential $v_{ext}({\bf r})$ at zero temperature. Based on the HK theorems \cite{HK}, 
the ground-state energy $E[\rho]$, a functional of the ground-state density $\rho({\bf r})$, can be written as
\begin{equation}
E[\rho] = F[\rho] + \int \rho({\bf r}) v_{ext}({\bf r}) d{\bf r},
\label{eq:dft7}
\end{equation}
where the universal functional 
\begin{equation}
F[\rho] = T[\rho] + V_{ee}[\rho], 
\label{eq:dft9qq}
\end{equation}
is the sum of the interacting kinetic energy $T[\rho]$ and the electron-electron repulsion energy $V_{ee}[\rho]$.

In KS-DFT \cite{HK,KS}, $F[\rho]$ is usefully partitioned as 
\begin{eqnarray}
F[\rho] &=& T_{s}[\rho] + E_{H}[\rho] + (T[\rho] + V_{ee}[\rho] - T_{s}[\rho] - E_{H}[\rho])  \nonumber  \\ 
           &=& T_{s}[\rho] + E_{H}[\rho] + E_{xc}[\rho].
\label{eq:dft9q}
\end{eqnarray}
Here $T_{s}[\rho]$ is the noninteracting kinetic energy at zero temperature,  
\begin{equation}
E_{H}[\rho] \equiv \frac{e^2}{2} \int\int \frac{\rho({\bf r})\rho({\bf r'})}{|{\bf r} - {\bf r'}|}d{\bf r}d{\bf r'}
\label{eq:hartree}
\end{equation}
is the Hartree energy, and $E_{xc}[\rho] \equiv (T[\rho] + V_{ee}[\rho] - T_{s}[\rho] - E_{H}[\rho])$ is the XC energy defined in KS-DFT. In KS-DFT, $T_{s}[\rho]$, the big unknown in terms of the density, is exactly treated 
by the use of KS orbitals. However, as previously discussed, the basic ansatz of KS-DFT (i.e.\ NI-PS $v_s$-representability of the given $\rho({\bf r})$) can be violated for SC systems, in which case 
even the exact KS-DFT will fail to convey reliably accurate results \cite{Perdew09}. 

To make progress, a different representation of the ground-state density is adopted in TAO-DFT, wherein $\rho({\bf r})$ is represented by the thermal equilibrium density of an auxiliary system of 
$N$ noninteracting electrons at a fictitious temperature $\theta$ in the presence of some local potential $v_{s}({\bf r})$. Aiming to achieve this representation, in contrast to the original KS partition, 
$F[\rho]$ is partitioned into the following set of terms: 
\begin{eqnarray}
F[\rho] &=& A_{s}^{\theta}[\rho] + E_{H}[\rho] + (T[\rho] + V_{ee}[\rho] - A_{s}^{\theta}[\rho] - E_{H}[\rho])  \nonumber  \\ 
           &=& A_{s}^{\theta}[\rho] + E_{H}[\rho] + (T[\rho] + V_{ee}[\rho] - T_{s}[\rho] - E_{H}[\rho]) + (T_{s}[\rho] - A_{s}^{\theta}[\rho]) \nonumber  \\ 
           &=& A_{s}^{\theta}[\rho] + E_{H}[\rho] + E_{xc}[\rho] + E_{\theta}[\rho].
\label{eq:dft9}
\end{eqnarray}
Here $T_{s}[\rho]$, $E_{H}[\rho]$, and $E_{xc}[\rho]$ are the same as those defined in KS-DFT, $A_{s}^{\theta}[\rho]$ is the noninteracting kinetic free energy at temperature $\theta$, and 
$E_{\theta}[\rho] \equiv T_{s}[\rho] - A_{s}^{\theta}[\rho] = A_{s}^{\theta=0}[\rho] - A_{s}^{\theta}[\rho]$ is the difference between the noninteracting kinetic free energy at zero temperature and that at temperature $\theta$. 

Substituting Eq.\ (\ref{eq:dft9}) into Eq.\ (\ref{eq:dft7}) and minimizing the $E[\rho]$ with respect to $\rho({\bf r})$ (subject to the constraint that the number of electrons be $N$), yields the following 
Euler equation for the ground-state density $\rho({\bf r})$, 
\begin{equation}
\mu = \frac{\delta A_{s}^{\theta}[\rho] }{\delta\rho({\bf r})} + v_{ext}({\bf r}) + e^2 \int \frac{\rho({\bf r'})}{|{\bf r} - {\bf r'}|}d{\bf r'} 
+ \frac{\delta E_{xc}[\rho]}{\delta \rho({\bf r})} + \frac{\delta E_{\theta}[\rho]}{\delta \rho({\bf r})},
\label{eq:dft8}
\end{equation}
where $\mu$ is the chemical potential of the system. 

To bypass the exact functional form of $A_{s}^{\theta}[\rho]$ (the big unknown in terms of the density) needed in Eq.\ (\ref{eq:dft8}), consider an auxiliary system of $N$ noninteracting electrons 
moving in a local potential $v_{s}({\bf r})$ at temperature $\theta$. Based on Mermin's theorems \cite{Mermin}, the grand-canonical potential $\Omega_{s}^{\theta}$ of this reference system, a functional 
of the thermal equilibrium density $\rho_{s}({\bf r})$, can be written as  
\begin{equation}
\Omega_{s}^{\theta}[\rho_{s}] = A_{s}^{\theta}[\rho_{s}] + \int \rho_{s}({\bf r}) [v_{s}({\bf r})-\mu_{s}] d{\bf r},
\label{eq:dft}
\end{equation}
where $\mu_{s}$ is the chemical potential of the reference system. 

Minimization of the $\Omega_{s}^{\theta}[\rho_{s}]$ with respect to $\rho_{s}({\bf r})$, gives the following Euler equation for the thermal equilibrium density $\rho_{s}({\bf r})$, 
\begin{equation}
\mu_{s} = \frac{\delta A_{s}^{\theta}[\rho_{s}] }{\delta\rho_{s}({\bf r})} + v_{s}({\bf r}). 
\label{eq:dft2}
\end{equation}
Comparing Eq.\ (\ref{eq:dft2}) with Eq.\ (\ref{eq:dft8}), shows that both minimizations have the same solution $\rho_{s}({\bf r})$=$\rho({\bf r})$, if we choose $v_{s}({\bf r})$ (up to a constant) as
\begin{equation}
v_{s}({\bf r}) = v_{ext}({\bf r}) + e^2 \int \frac{\rho({\bf r'})}{|{\bf r} - {\bf r'}|}d{\bf r'}  + \frac{\delta E_{xc}[\rho]}{\delta \rho({\bf r})} 
+ \frac{\delta E_{\theta}[\rho]}{\delta \rho({\bf r})}. 
\label{eq:dft10}
\end{equation}

Alternatively, as $A_{s}^{\theta}[\rho]$ can be expressed exactly in terms of orbitals and occupation numbers (see below), Eq.\ (\ref{eq:dft2}) can also be handled, in an exact manner, by solving the 
one-electron Schr\"odinger equations for the potential $v_{s}({\bf r})$, given by
\begin{equation}
\{-\frac{\hbar^2}{2 m_e} {\bf \nabla}^2 \ + \ v_{s}({\bf r})\} \psi_{i}({\bf r}) = \epsilon_{i} \psi_{i}({\bf r}), 
\label{eq:dft3}
\end{equation}
and construct 
\begin{equation}
\rho({\bf r}) = \rho_{s}({\bf r}) = \sum_{i=1}^{\infty} f_{i} |\psi_{i}({\bf r}) |^{2},
\label{eq:dft4}
\end{equation}
where the occupation number $f_{i}$ is the Fermi-Dirac function
\begin{equation}
f_{i} = \{1+\text{exp}[( \epsilon_{i} - \mu)/ \theta] \}^{-1},
\label{eq:dft5}
\end{equation}
and the chemical potential $\mu$ is chosen to conserve the number of electrons $N$, 
\begin{equation}
\sum_{i=1}^{\infty} \{1+\text{exp}[( \epsilon_{i} - \mu)/ \theta] \}^{-1} = N. 
\label{eq:dft6}
\end{equation}
The formulation of TAO-DFT leads to a set of self-consistent equations, Eqs.\ (\ref{eq:dft10}), (\ref{eq:dft3}), (\ref{eq:dft4}), (\ref{eq:dft5}), and (\ref{eq:dft6}). 

To obtain a self-consistent ground-state density in TAO-DFT: (i) Choose a trial $\rho({\bf r})$ to construct $v_{s}({\bf r})$ by Eq.\ (\ref{eq:dft10}); (ii) solve Eq.\ (\ref{eq:dft3}), which 
gives $\{ \epsilon_{i}, \psi_{i}({\bf r}) \}$; (iii) find $\mu$ by solving Eq.\ (\ref{eq:dft6}); (iv) determine $\{ f_{i} \}$ by Eq.\ (\ref{eq:dft5}) and new $\rho({\bf r})$ by Eq.\ (\ref{eq:dft4}).\ 
This process is coupled with Eq.\ (\ref{eq:dft10}) to achieve self-consistency. When converged, the entropy functional reads
\begin{equation} 
S_{s}^{\theta}[\{ f_{i}\}] = -k_{B} \sum_{i=1}^{\infty} [f_{i}\ \text{ln}(f_{i}) + (1-f_{i})\ \text{ln}(1-f_{i})].   
\label{eq:dft11b}
\end{equation} 
The exact noninteracting kinetic free energy $A_{s}^{\theta}$ at the fictitious temperature $\theta$, can be expressed in terms of $\{ f_{i}, \psi_{i} \}$:
\begin{equation}
A_{s}^{\theta}[\{ f_{i}, \psi_{i} \}] = T_{s}^{\theta}[\{ f_{i}, \psi_{i} \}] - \frac{\theta}{k_{B}} S_{s}^{\theta}[\{ f_{i}\}], 
\label{eq:dft11c}
\end{equation}
which is the sum of the kinetic energy 
\begin{eqnarray}
T_{s}^{\theta}[\{ f_{i}, \psi_{i} \}] &=& -\frac{\hbar^2}{2 m_e} \sum_{i=1}^{\infty} f_{i} \int \psi_{i}^{*}({\bf r}){\bf \nabla}^2\psi_{i}({\bf r}) d{\bf r} \nonumber  \\ 
                                                      &=& \sum_{i=1}^{\infty} f_{i}  \epsilon_{i} - \int \rho({\bf r}) v_{s}({\bf r}) d{\bf r}
\label{eq:dft11a}
\end{eqnarray}
and entropy contribution  
\begin{equation} 
- \frac{\theta}{k_{B}} S_{s}^{\theta}[\{ f_{i}\}] =  \theta \sum_{i=1}^{\infty} [f_{i}\ \text{ln}(f_{i}) + (1-f_{i})\ \text{ln}(1-f_{i})]   
\label{eq:dft11b}
\end{equation} 
of noninteracting electrons at temperature $\theta$.  
Based on Eqs.\ (\ref{eq:dft7}) and (\ref{eq:dft9}), the total ground-state energy $E[\rho]$ can be evaluated by
\begin{equation} 
E[\rho] = A_{s}^{\theta}[\{ f_{i}, \psi_{i} \}]  + \int \rho({\bf r}) v_{ext}({\bf r}) d{\bf r} + E_{H}[\rho] + E_{xc}[\rho] + E_{\theta}[\rho].
\label{eq:dft12}
\end{equation} 

To sum up, in TAO-DFT, the partition of $F[\rho]$ in Eq.\ (\ref{eq:dft9}) and the exact treatment of $A_{s}^{\theta}[\rho]$ in Eq.\ (\ref{eq:dft11c}), are shown to yield a set of self-consistent equations 
for the NI-TE $v_s$-representable $\rho({\bf r})$ in Eq.\ (\ref{eq:dft4}). Note that these equations resemble the finite-temperature KS equations \cite{KS,Mermin}, so the implementation of TAO-DFT 
can be easily achieved using existing FT-DFT codes with a slight modification (i.e.\ replacing the XC free energy in FT-DFT to the sum of $E_{xc}[\rho]$ and $E_{\theta}[\rho]$ in TAO-DFT). 
Hence, the computational cost of TAO-DFT is similar to that of KS-DFT or FT-DFT. Similar to FT-DFT \cite{KS,Mermin}, due to the explicit inclusion of Fermi-Dirac occupation function in TAO-DFT, 
the entropy contribution ($- \frac{\theta}{k_{B}} S_{s}^{\theta}[\{ f_{i}\}]$) in Eq.\ (\ref{eq:dft11b}) is essential to make the total ground-state energy functional $E[\rho]$ variational \cite{Weinert} 
(e.g.\ making the nuclear gradients of $E[\rho]$ equal to the Hellmann-Feynman forces \cite{HF}).

\subsection{Spin-Polarized Formalism}

For a system with $N_{\alpha}$ up-spin electrons and $N_{\beta}$ down-spin electrons, the standard computational approach is the spin-polarized (spin-unrestricted) formalism, wherein 
the fundamental variables are the up-spin density $\rho_{\alpha}({\bf r})$ and down-spin density $\rho_{\beta}({\bf r})$ of the ground-state density 
\begin{equation}
\rho({\bf r}) = \rho_{\alpha}({\bf r}) + \rho_{\beta}({\bf r}) = \sum_{\sigma=\alpha,\beta} \rho_{\sigma}({\bf r}).  
\label{eq:sdft4a}
\end{equation}
In analogy to the two-Fermi-level picture of spin-polarized KS-DFT \cite{SDFT,SDFTYang}, spin-polarized TAO-DFT can also be formulated with the two-chemical-potential picture, wherein two noninteracting 
auxiliary systems at the same fictitious temperature $\theta$ are adopted: one described by the spin function $\alpha$ and the other by function $\beta$, with the corresponding thermal equilibrium density 
distributions $\rho_{s,\alpha}({\bf r})$ and $\rho_{s,\beta}({\bf r})$ exactly equal to $\rho_{\alpha}({\bf r})$ and $\rho_{\beta}({\bf r})$, respectively, in the original interacting system at zero temperature. Similar to 
the previous derivations (but using the spin-polarized extensions of the HK theorems \cite{SDFT,SDFTYang} and the Mermin theorems \cite{SFTDFT,SFTDFT2} for the physical and auxiliary systems, respectively), 
one-electron Schr\"odinger equations for electrons with $\sigma$-spin ($\sigma$ = $\alpha$ or $\beta$), can be obtained as follows ($i$ runs for the orbital index):  
\begin{equation}
\{-\frac{\hbar^2}{2 m_e} {\bf \nabla}^2 \ + \ v_{s,\sigma}({\bf r})\} \psi_{i,\sigma}({\bf r}) = \epsilon_{i,\sigma} \psi_{i,\sigma}({\bf r}), 
\label{eq:sdft3}
\end{equation}
with a local potential   
\begin{equation}
v_{s,\sigma}({\bf r}) = v_{ext}({\bf r}) + e^2 \int \frac{\rho({\bf r'})}{|{\bf r} - {\bf r'}|}d{\bf r'}  + \frac{\delta E_{xc}[\rho_{\alpha},\rho_{\beta}]}{\delta \rho_{\sigma}({\bf r})} 
+ \frac{\delta E_{\theta}[\rho_{\alpha},\rho_{\beta}]}{\delta \rho_{\sigma}({\bf r})}. 
\label{eq:sdft10}
\end{equation}
Here $E_{xc}[\rho_{\alpha},\rho_{\beta}]$ is the same as the XC energy defined in spin-polarized KS-DFT \cite{SDFT,SDFTYang}, and 
$E_{\theta}[\rho_{\alpha},\rho_{\beta}] \equiv T_{s}[\rho_{\alpha},\rho_{\beta}] - A_{s}^{\theta}[\rho_{\alpha},\rho_{\beta}] = A_{s}^{\theta=0}[\rho_{\alpha},\rho_{\beta}] - A_{s}^{\theta}[\rho_{\alpha},\rho_{\beta}]$ is 
the spin-polarized version of $E_{\theta}[\rho]$. The $\sigma$-spin density can be constructed by 
\begin{equation}
\rho_{\sigma}({\bf r}) = \sum_{i=1}^{\infty} f_{i,\sigma} |\psi_{i,\sigma}({\bf r}) |^{2},
\label{eq:sdft4}
\end{equation}
where the occupation number $f_{i,\sigma}$ is the Fermi-Dirac function
\begin{equation}
f_{i,\sigma} = \{1+\text{exp}[( \epsilon_{i,\sigma} - \mu_{\sigma})/ \theta] \}^{-1},
\label{eq:sdft5}
\end{equation}
and a chemical potential $\mu_{\sigma}$ is chosen to conserve the number of $\sigma$-spin electrons $N_{\sigma}$, 
\begin{equation}
\sum_{i=1}^{\infty} \{1+\text{exp}[( \epsilon_{i,\sigma} - \mu_{\sigma})/ \theta] \}^{-1} = N_{\sigma}. 
\label{eq:sdft6}
\end{equation}
The formulation of spin-polarized TAO-DFT yields two sets (one for each spin function) of self-consistent equations, Eqs.\ (\ref{eq:sdft3}), (\ref{eq:sdft10}), (\ref{eq:sdft4}), (\ref{eq:sdft5}), and (\ref{eq:sdft6}), for 
$\rho_{\alpha}({\bf r})$ and $\rho_{\beta}({\bf r})$, respectively, which are coupled with $\rho({\bf r})$ by Eq.\ (\ref{eq:sdft4a}). 

To obtain self-consistent spin densities (and the ground-state density) in spin-polarized TAO-DFT: 
(i) Choose trial spin densities $\rho_{\alpha}({\bf r})$ and $\rho_{\beta}({\bf r})$ to compute the ground-state density $\rho({\bf r})$ by Eq.\ (\ref{eq:sdft4a}); 
(ii) for the $\sigma$-spin ($\sigma$ = $\alpha$ or $\beta$) electrons, construct $v_{s,\sigma}({\bf r})$ by Eq.\ (\ref{eq:sdft10}); 
(iii) solve Eq.\ (\ref{eq:sdft3}), which gives $\{ \epsilon_{i,\sigma}, \psi_{i,\sigma}({\bf r}) \}$; 
(iv) find $\mu_{\sigma}$ by solving Eq.\ (\ref{eq:sdft6}); 
(v) determine $\{ f_{i,\sigma} \}$ by Eq.\ (\ref{eq:sdft5}) and new $\rho_{\sigma}({\bf r})$ by Eq.\ (\ref{eq:sdft4}). 
This process is coupled with Eq.\ (\ref{eq:sdft4a}) to achieve self-consistency. 
When converged, $A_{s,\sigma}^{\theta}$, the sum of the kinetic energy and entropy contribution of noninteracting $\sigma$-spin electrons at the fictitious temperature $\theta$, is given by 
\begin{eqnarray}
A_{s,\sigma}^{\theta}[\{ f_{i,\sigma}, \psi_{i,\sigma} \}]  &=&-\frac{\hbar^2}{2 m_e} \sum_{i=1}^{\infty} f_{i,\sigma} \int \psi_{i,\sigma}^{*}({\bf r}){\bf \nabla}^2\psi_{i,\sigma}({\bf r}) d{\bf r} 
                           +\ \theta \sum_{i=1}^{\infty} [f_{i,\sigma}\ \text{ln}(f_{i,\sigma}) + (1-f_{i,\sigma})\ \text{ln}(1-f_{i,\sigma})]  \nonumber  \\ 
                    &=& \sum_{i=1}^{\infty} \{ f_{i,\sigma}  \epsilon_{i,\sigma}  + \theta \ [f_{i,\sigma}\ \text{ln}(f_{i,\sigma}) + (1-f_{i,\sigma})\ \text{ln}(1-f_{i,\sigma}) ]\}
                         - \int \rho_{\sigma}({\bf r}) v_{s,\sigma}({\bf r}) d{\bf r},
\label{eq:sdft11}
\end{eqnarray}
and the total ground-state energy $E[\rho_{\alpha},\rho_{\beta}]$ in spin-polarized TAO-DFT is evaluated by
\begin{equation}
E[\rho_{\alpha},\rho_{\beta}] = \sum_{\sigma=\alpha,\beta} A_{s,\sigma}^{\theta}[\{ f_{i,\sigma}, \psi_{i,\sigma} \}]  + \int \rho({\bf r}) v_{ext}({\bf r}) d{\bf r} + E_{H}[\rho] + E_{xc}[\rho_{\alpha},\rho_{\beta}] + E_{\theta}[\rho_{\alpha},\rho_{\beta}].
\label{eq:sdft12}
\end{equation}

Spin-unpolarized (spin-restricted) TAO-DFT can be formulated by imposing the constraints of $\psi_{i,\alpha}({\bf r})$ = $\psi_{i,\beta}({\bf r})$ and $f_{i,\alpha}$ = $f_{i,\beta}$ to spin-polarized (spin-unrestricted) TAO-DFT.

\subsection{Analytical Nuclear Gradients}

The analytical computation of nuclear gradients is crucial for the efficient optimization of molecular geometries. In light of the similarity of TAO-DFT and FT-DFT \cite{KS,Mermin}, analytical nuclear gradients for TAO-DFT 
can be easily obtained from those for FT-DFT with a slight modification (mentioned previously). Therefore, the computational cost of the analytical nuclear gradients for TAO-DFT is similar to that for KS-DFT or FT-DFT. For 
nonorthogonal atomic-orbital representations (e.g.\ Gaussian-type orbitals), the generalized Pulay force for a noninteracting thermal ensemble \cite{grad} should be included to add the effect of the basis-set dependent response 
to the analytical nuclear gradients for TAO-DFT.

\subsection{Local Density Approximation}

In TAO-DFT, as the exact $E_{xc}[\rho]$ and $E_{\theta}[\rho]$, in terms of the ground-state density $\rho({\bf r})$, remain unknown, DFAs for both of them (denoted as TAO-DFAs) are needed for practical applications. 
The performance of TAO-DFAs depends on the accuracy of DFAs and the choice of the fictitious temperature $\theta$. In this work, we adopt the LDA (the simplest DFA) for both the $E_{xc}[\rho]$ and $E_{\theta}[\rho]$ 
in TAO-DFT (denoted as TAO-LDA). As TAO-LDA is exact for a uniform electron gas (UEG), it provides a good starting point for more accurate and sophisticated TAO-DFAs. Besides, TAO-LDA is readily available, 
as $E_{xc}^{\text {LDA}}[\rho]$ can be directly obtained from that of KS-LDA \cite{LDAX,LDAC}, and $E_{\theta}^{\text {LDA}}[\rho]$ can be obtained with the knowledge of $A_{s}^{\text{LDA}, \theta}[\rho]$ as follows: 
\begin{equation}
E_{\theta}^{\text {LDA}}[\rho] \equiv A_{s}^{\text{LDA}, \theta=0}[\rho] - A_{s}^{\text{LDA}, \theta}[\rho].
\label{eq:ftdft1}
\end{equation}
Here, Perrot's parametrization of $A_{s}^{\text{LDA}, \theta}[\rho]$ (in its spin-unpolarized form) \cite{As} is adopted to obtain $E_{\theta}^{\text {LDA}}[\rho]$ (in its spin-unpolarized form) by Eq.\ (\ref{eq:ftdft1}). 
For completeness of this work, $A_{s}^{\text{LDA}, \theta}[\rho]$ (after correcting some typos in Ref.\ \cite{As}) is explicitly shown here (in atomic units), 
\begin{equation}
A_{s}^{\text{LDA}, \theta}[\rho] = \int a_{s}^{\text{LDA}, \theta}({\bf r}) d{\bf r}, 
\label{eq:ftdft1z}
\end{equation}
where $a_{s}^{\text{LDA}, \theta}({\bf r}) \equiv \theta \rho({\bf r}) f(y)$ and $y \equiv (\pi^2/\sqrt{2})\theta^{-3/2}\rho({\bf r})$. The function $f(y)$ was parametrized separately for the two regions 
$y \le y_{0}$ and $y \ge y_{0}$ ($y_{0} \equiv \frac{3 \pi}{4 \sqrt{2}}$) \cite{As}: 
\begin{eqnarray}
f(y) &=& \text{ln}y - 0.8791880215 + 0.1989718742 y + 0.1068697043 \times 10^{-2} y^{2} \nonumber  \\  
&-&  0.8812685726 \times 10^{-2} y^{3} + 0.1272183027 \times 10^{-1} y^{4} - 0.9772758583 \times 10^{-2} y^{5} \nonumber  \\
&+&  0.3820630477 \times 10^{-2} y^{6} - 0.5971217041 \times 10^{-3} y^{7},\ \text{for}\ y \le y_{0}, 
\label{eq:ftdft1z2}
\end{eqnarray}
\begin{eqnarray}
f(y) &=& 0.7862224183 u - 0.1882979454 \times 10^{1} u^{-1} + 0.5321952681 u^{-3} \nonumber  \\  
&+& 0.2304457955 \times 10^{1} u^{-5} - 0.1614280772 \times 10^{2} u^{-7} + 0.5228431386 \times 10^{2} u^{-9} \nonumber  \\  
&-& 0.9592645619 \times 10^{2} u^{-11} + 0.9462230172 \times 10^{2} u^{-13} \nonumber  \\  
&-& 0.3893753937 \times 10^{2} u^{-15},\ \text{for}\ y \ge y_{0}\ \text{with}\ u \equiv y^{2/3}.  
\label{eq:ftdft1z3}
\end{eqnarray}
The $\theta=0$ case, $A_{s}^{\text{LDA}, \theta=0}[\rho]$, is the same as the Thomas-Fermi kinetic energy density functional \cite{Parr,Kohanoff}, 
\begin{equation}
A_{s}^{\text{LDA}, \theta=0}[\rho] = C_{F} \int \rho^{5/3}({\bf r}) d{\bf r}, 
\label{eq:ftdft1za}
\end{equation}
where $C_{F} = \frac{3}{10}(3 \pi^2)^{2/3}$. 

For spin-polarized (spin-unrestricted) TAO-LDA, the corresponding spin-polarized forms, $E_{xc}^{\text {LDA}}[\rho_{\alpha},\rho_{\beta}]$ (available from that of spin-polarized 
KS-LDA \cite{LDAX,LDAC}) and $E_{\theta}^{\text {LDA}}[\rho_{\alpha},\rho_{\beta}]$, should be adopted. From the spin-scaling relation of $A_{s}^{\theta}[\rho_{\alpha},\rho_{\beta}]$ (same as that 
of $T_{s}[\rho_{\alpha},\rho_{\beta}]$ \cite{spin-scaling}), $E_{\theta}^{\text {LDA}}[\rho_{\alpha},\rho_{\beta}]$ can be conveniently expressed by its spin-unpolarized form $E_{\theta}^{\text {LDA}}[\rho]$, 
\begin{eqnarray}
E_{\theta}^{\text {LDA}}[\rho_{\alpha},\rho_{\beta}] &\equiv& A_{s}^{\text{LDA}, \theta=0}[\rho_{\alpha},\rho_{\beta}] - A_{s}^{\text{LDA}, \theta}[\rho_{\alpha},\rho_{\beta}] \nonumber  \\ 
&=& \frac{1}{2} (A_{s}^{\text{LDA}, \theta=0}[2\rho_{\alpha}] + A_{s}^{\text{LDA}, \theta=0}[2\rho_{\beta}]) - \frac{1}{2} (A_{s}^{\text{LDA}, \theta}[2\rho_{\alpha}] + A_{s}^{\text{LDA}, \theta}[2\rho_{\beta}]) \nonumber  \\ 
&=&\frac{1}{2} \{(A_{s}^{\text{LDA}, \theta=0}[2\rho_{\alpha}] - A_{s}^{\text{LDA}, \theta}[2\rho_{\alpha}]) + (A_{s}^{\text{LDA}, \theta=0}[2\rho_{\beta}] - A_{s}^{\text{LDA}, \theta}[2\rho_{\beta}])\} \nonumber  \\  
&=& \frac{1}{2} \{E_{\theta}^{\text {LDA}}[2\rho_{\alpha}] + E_{\theta}^{\text {LDA}}[2\rho_{\beta}]\}.  
\label{eq:ftdft2}
\end{eqnarray}

\subsection{Strong Static Correlation from TAO-LDA}

The ground-state density $\rho({\bf r})$ of a strongly correlated system containing a sufficiently large number of electrons, can be represented by Eq.\ (\ref{eq:rhoci}) (with the exact NOs $\{\chi_{i}({\bf r})\}$ 
and NOONs $\{n_{i}\}$). Assume that the $\rho({\bf r})$ can also be represented by Eq.\ (\ref{eq:dft4}) (with the orbitals $\{\psi_{i}({\bf r})\}$ and their occupation numbers $\{f_{i}\}$ from the exact TAO-DFT). Note that 
for such a NI-TE $v_s$-representable $\rho({\bf r})$, its "internal variables" $\{f_{i}\}$ and $\{\psi_{i}({\bf r})\}$ in Eq.\ (\ref{eq:dft4}), can still be tuned by changing the fictitious temperature $\theta$. If a $\theta$ is chosen 
so that $\{n_{i}\}$ $\approx$ $\{f_{i}\}$ (in the sense of statistical average, as mentioned previously), we have $\{\chi_{i}({\bf r})\}$ $\approx$ $\{\psi_{i}({\bf r})\}$ (as both Eq.\ (\ref{eq:rhoci}) and Eq.\ (\ref{eq:dft4}) 
represent the same $\rho({\bf r})$). In fact, the NI-TE $v_s$-representability of $\rho({\bf r})$ is likely to be fulfilled for this $\theta$, due to the similarity of Eq.\ (\ref{eq:rhoci}) and Eq.\ (\ref{eq:dft4}). 

Consequently, the exact kinetic energy of {\it interacting} electrons $T[\rho]$ can be properly simulated by $T_{s}^{\theta}[\{ f_{i}, \psi_{i} \}]$ 
(as appeared in $A_{s}^{\theta}[\{ f_{i}, \psi_{i} \}]$ = $T_{s}^{\theta}[\{ f_{i}, \psi_{i} \}] - \frac{\theta}{k_{B}} S_{s}^{\theta}[\{ f_{i}\}]$), namely,  
\begin{equation}
T[\rho] = T[\{ n_{i}, \chi_{i} \}] \approx T_{s}^{\theta}[\{ f_{i}, \psi_{i} \}], 
\label{eq:dft28}
\end{equation}
due to their similar expressions \cite{Jensen}, while the electron-electron repulsion energy $V_{ee}[\rho] = F[\rho] - T [\rho]$ (see Eqs.\ (\ref{eq:dft9qq}) and (\ref{eq:dft9})) is given  by, 
\begin{equation}
V_{ee}[\rho] \approx E_{H}[\rho] + E_{xc}[\rho] + E_{\theta}[\rho] - \frac{\theta}{k_{B}} S_{s}^{\theta}[\{ f_{i}\}]. 
\label{eq:dft29e}
\end{equation}
On the right-hand side of Eq.\ (\ref{eq:dft29e}), the first term is the Hartree energy, and the sum of the remaining terms ($E_{xc}[\rho] + E_{\theta}[\rho] - \frac{\theta}{k_{B}} S_{s}^{\theta}[\{ f_{i}\}]$) should properly 
describe the XC energy defined in the exact wave function theory. 

Here, we explain how strong static correlation is described by TAO-LDA, with arguments similar to the above. 
Suppose that the exact $\rho({\bf r})$ in Eq.\ (\ref{eq:rhoci}) can be reasonably represented by Eq.\ (\ref{eq:dft4}) (with the orbitals $\{\psi_{i}({\bf r})\}$ 
and their occupation numbers $\{f_{i}\}$ from TAO-LDA). 
When applying the above arguments for TAO-LDA (i.e.\ choosing a $\theta$ so that $\{n_{i}\}$ $\approx$ $\{f_{i}\}$, which gives $\{\chi_{i}({\bf r})\}$ $\approx$ $\{\psi_{i}({\bf r})\}$), $T[\rho]$ can still be properly simulated 
by $T_{s}^{\theta}[\{ f_{i}, \psi_{i} \}]$ (see Eq.\ (\ref{eq:dft28})), while $V_{ee}[\rho]$ is only approximated by, 
\begin{equation}
V_{ee}[\rho] \approx E_{H}[\rho] + E_{xc}^{\text {LDA}}[\rho] + E_{\theta}^{\text {LDA}}[\rho] - \frac{\theta}{k_{B}} S_{s}^{\theta}[\{ f_{i}\}]. 
\label{eq:dft29}
\end{equation}
On the right-hand side of Eq.\ (\ref{eq:dft29}), the first two terms ($E_{H}[\rho]$ and $E_{xc}^{\text {LDA}}[\rho]$) are the same as those defined in KS-LDA, the third term $E_{\theta}^{\text {LDA}}[\rho]$ locally accounts for 
the difference between the exact $T_{s}$ and $A_{s}^{\theta}$ (at the LDA level), and the last term is the entropy contribution 
($- \frac{\theta}{k_{B}} S_{s}^{\theta}[\{ f_{i}\}]$ $\approx$ $- \frac{\theta}{k_{B}} S_{s}^{\theta}[\{ n_{i}\}]$ = $\theta \sum_{i=1}^{\infty} [n_{i}\ \text{ln}(n_{i}) + (1-n_{i})\ \text{ln}(1-n_{i})]$) (see Eq.\ (\ref{eq:dft11b})). 

Due to their local treatment, $E_{xc}^{\text {LDA}}[\rho]$ and $E_{\theta}^{\text {LDA}}[\rho]$ are not expected to properly describe nonlocal XC effects (e.g.\ long-range dynamical correlation, strong static correlation). 
However, as the entropy contribution is a fully nonlocal density functional ($\{ f_{i}\}$ are implicit density functionals), it may describe nonlocal correlation effects. 
There is certainly a close relationship between the entropy (defined by the NOONs $\{n_{i}\}$) and correlation energy of a system. 
A famous example is given by the Collins conjecture \cite{Collins} that the correlation energy of a system is proportional to the Jaynes (information) entropy 
$S_{\text{Jaynes}}[\{ n_{i}\}] = - \sum_{i=1}^{\infty} n_{i}\ \text{ln}(n_{i})$ \cite{Jaynes}. Interestingly, the entropy contribution in Eq.\ (\ref{eq:dft29}) is proportional to the Gibbs (thermodynamic) 
entropy ($S_{s}^{\theta}[\{ f_{i}\}]$ $\approx$ $S_{s}^{\theta}[\{ n_{i}\}]$ = $- k_{B} \sum_{i=1}^{\infty} [n_{i}\ \text{ln}(n_{i}) + (1-n_{i})\ \text{ln}(1-n_{i})]$), with the constant of proportionality being explicitly given 
by ($- \frac{\theta}{k_{B}}$). Note that the similarity of information entropy and thermodynamic entropy in a many-body quantum system (with strong interactions) has been shown in Ref.\ \cite{Horoi}, based on statistical arguments. 

As the entropy contribution $(- \frac{\theta}{k_{B}} S_{s}^{\theta}[\{f_{i}\}])$ in Eq.\ (\ref{eq:dft29}) essentially provides no contributions for a single-reference system ($\{f_{i}\}$ $\approx$ $\{n_{i}\}$ are close to either 0 or 1), 
and significantly lowers the total energy of a multi-reference system ($\{f_{i}\}$ $\approx$ $\{n_{i}\}$ are fractional for active orbitals, and are close to either 0 or 1 for others), we expect that this term (absent in KS-LDA) play 
a crucial role in simulating strong static correlation (rather than dynamical correlation).

\section{Numerical Investigations of an Optimal $\theta$ Value}

The fictitious (reference) temperature $\theta$ for TAO-DFT, controlling the orbital occupation numbers $\{ f_{i}\}$, is closely related to the strength of static correlation. At the LDA level, an immediate question is how the $\theta$ 
for the resulting TAO-LDA should be chosen. As previously argued, the entropy contribution, $- \frac{\theta}{k_{B}} S_{s}^{\theta}[\{ f_{i}\}] = \theta \sum_{i=1}^{\infty} [f_{i}\ \text{ln}(f_{i}) + (1-f_{i})\ \text{ln}(1-f_{i})]$, 
can be responsible for strong static correlation effects, especially when the $\{ f_{i}\}$ (tunable by the $\theta$) properly simulate the exact NOONs $\{n_{i}\}$. To numerically investigate this conjecture, 
the performance of TAO-LDA (with $\theta$ = 0, 1, 3, 5, 7, 10, 15, 20, 30, and 40 mHartree) is examined for both single-reference systems (reaction energies and equilibrium geometries) and multi-reference systems 
(dissociation of H$_2$ and N$_2$, twisted ethylene, and singlet-triplet energy gaps of linear acenes). The limiting case where $\theta=0$ for TAO-LDA is especially interesting, as this reduces to KS-LDA. Therefore, 
it is important to know how well KS-LDA performs here to assess the significance of the extra parameter $\theta$ for TAO-LDA. 

All calculations are performed with a development version of \textsf{Q-Chem 3.2} \cite{QChem}. The error for each entry is defined as (error = theoretical value $-$ reference value). The notation used for characterizing 
statistical errors is as follows: mean signed errors (MSEs), mean absolute errors (MAEs), root-mean-square (rms) errors, maximum negative errors (Max($-$)), and maximum positive errors (Max(+)). Results are computed 
using the 6-311++G(3df,3pd) basis set, unless noted otherwise.

\subsection{Single-Reference Systems}

\subsubsection{Reaction Energies}

The accurate prediction of reaction energies is usually one of the major criteria in the assessment of the performance of electronic structure methods. The reaction energies of 30 chemical reactions 
(a test set described in Ref.\ \cite{wB97X}) are used to examine the performance of TAO-LDA. As shown in Table \ref{table:reall}, TAO-LDA (with a $\theta$ smaller than 10 mHartree) has similar performance to 
KS-LDA \cite{supp}. This is unsurprising, as these systems do not have much static correlation, the exact NOONs should be close to either 0 or 1, which can be well simulated by the orbital occupation numbers of 
TAO-LDA (with a sufficiently small $\theta$). Consequently, $T_{s}^{\theta}[\{ f_{i}, \psi_{i} \}]$ (see Eq.\ (\ref{eq:dft28})) is close to $T_{s}^{\theta=0}[\{ \psi_{i} \}]$ (KS kinetic energy), and 
$E_{\theta}^{\text {LDA}}[\rho]$ (see Eq.\ (\ref{eq:ftdft1})) 
and the entropy contribution ($- \frac{\theta}{k_{B}} S_{s}^{\theta}[\{ f_{i}\}] = \theta \sum_{i=1}^{\infty} [f_{i}\ \text{ln}(f_{i}) + (1-f_{i})\ \text{ln}(1-f_{i})]$) have insignificant contributions to the total energy, 
relative to $E_{xc}^{\text {LDA}}[\rho]$ (see Eq.\ (\ref{eq:dft29})).

\subsubsection{Equilibrium Geometries}

Geometry optimizations for TAO-LDA are performed on the equilibrium experimental test set (EXTS) \cite{EXTS}, consisting of 166 symmetry unique experimental bond lengths for small to medium size molecules. 
As the ground states of these molecules near their equilibrium geometries can be well described by single-reference wave functions, TAO-LDA (with a $\theta$ smaller than 10 mHartree) is also found to 
perform similarly to KS-LDA \cite{supp}, as shown in Table \ref{table:EXTS}.

\subsection{Multi-Reference Systems}

\subsubsection{Dissociation of H$_2$ and N$_2$}

\begin{figure}
\caption{\label{h2t} Potential energy curves (in total energy) for the ground state of H$_2$, calculated by both the spin-restricted and spin-unrestricted formalisms of the HF theory and 
KS-DFT (with LDA and B3LYP functionals). The exact potential energy curve is calculated by the CCSD theory.}
\includegraphics[scale=1.0]{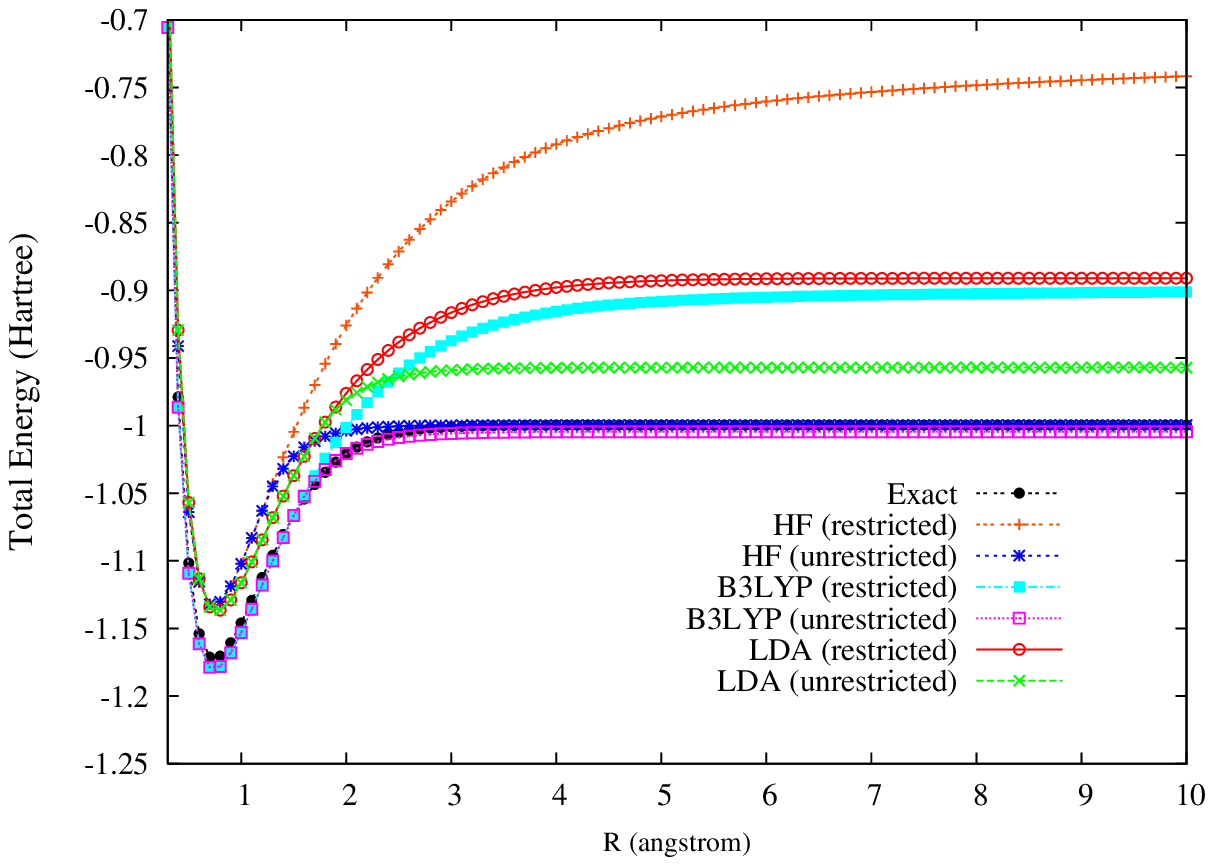}
\end{figure}

\begin{figure}
\caption{\label{h2t1} Potential energy curves (in relative energy) for the ground state of H$_2$, calculated by the spin-restricted HF theory and KS-DFT (with various XC functionals). The exact 
potential energy curve is calculated by the CCSD theory. The zeros of energy are set at the respective spin-unrestricted dissociation limits.}
\includegraphics[scale=1.0]{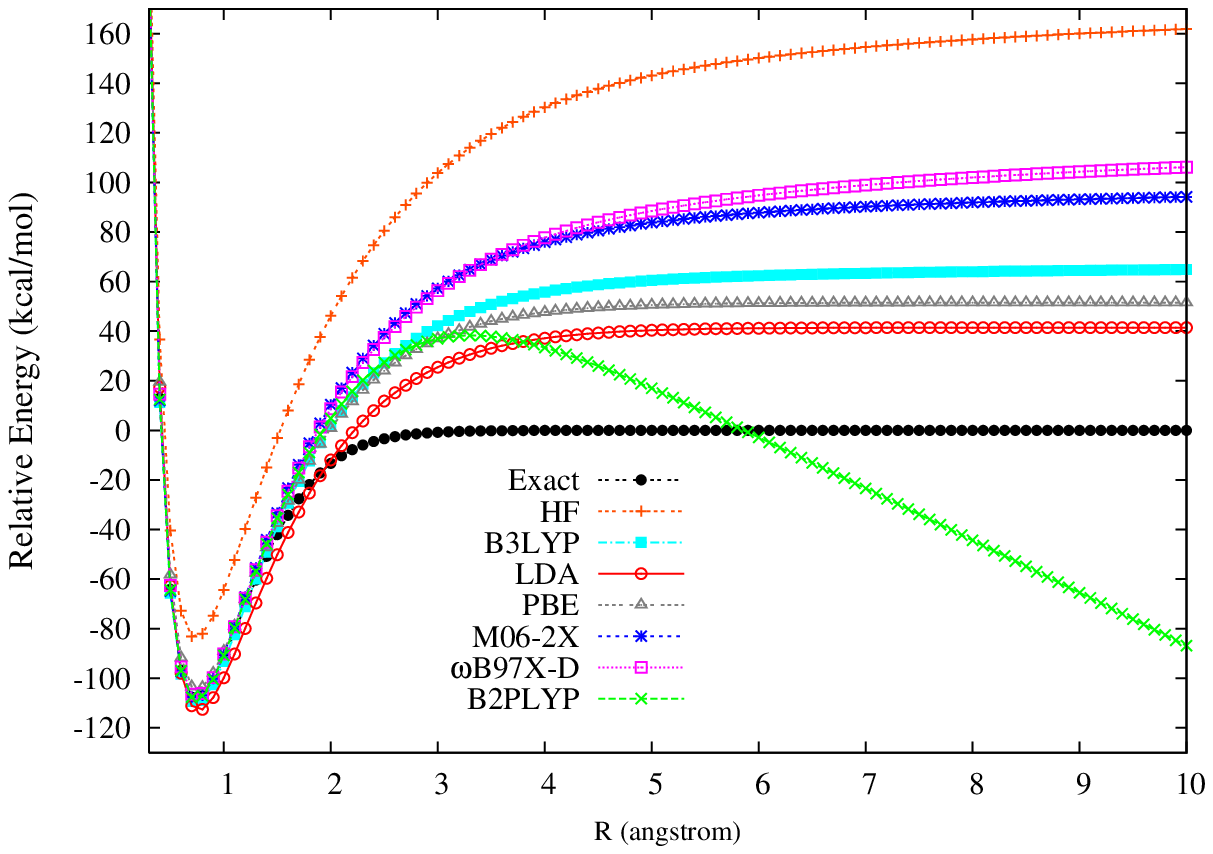}
\end{figure}

H$_2$ dissociation, a single-bond breaking system, is particularly challenging for KS-DFT. Fig.\ (\ref{h2t}) shows the potential energy curves (in total energy) for the ground state of H$_2$, 
calculated by both the spin-restricted and spin-unrestricted formalisms of the HF theory and KS-DFT (with LDA \cite{LDAX,LDAC} and B3LYP \cite{hybrid,B3LYP} functionals), where the exact potential energy curve 
is calculated by the CCSD theory (coupled-cluster theory with iterative singles and doubles) \cite{CCSD}. Due to the symmetry constraint, the spin-restricted and spin-unrestricted potential energy curves, 
calculated by the exact theory, should be the same. Therefore, the difference between the dissociation limits of the spin-restricted and spin-unrestricted potential energy curves, can be used as 
a quantitative measure of SCEs of approximate methods \cite{SciYang,SCE}. Spin-restricted KS-DFT yields the proper spin symmetry and spin densities, but has much too high total energy 
(leading to a noticeable SCE \cite{SciYang,SCE}), due to the lack of strong static correlation. On the other hand, spin-unrestricted KS-DFT artificially breaks the correct space- and spin-symmetries to simulate 
strong static correlation, yielding a reasonable energy but wrong spin densities \cite{SciYang}. Similar results are also found for the HF theory. Among the three approximate methods, HF has the largest SCE 
due to the complete neglect of static (and also dynamical) correlation. The SCE of LDA is smaller than that of B3LYP or HF. In Fig.\ (\ref{h2t1}), the potential energy curves (in relative energy) for the 
ground state of H$_2$, calculated by the spin-restricted HF theory and KS-DFT (with LDA \cite{LDAX,LDAC}, PBE \cite{PBE}, B3LYP \cite{hybrid,B3LYP}, M06-2X \cite{M06-2X}, $\omega$B97X-D \cite{wB97X-D}, 
and B2PLYP \cite{B2PLYP} functionals), are presented for comparisons, where the zeros of energy are set at the respective spin-unrestricted dissociation limits. As can be seen, LDA still has the smallest SCE, 
when compared with other approximate methods. Popular hybrid functionals (B3LYP, M06-2X, and $\omega$B97X-D) perform very well near the equilibrium geometry (dominated by single-reference character), 
but fail drastically at the larger $R$ (dominated by multi-reference character). B2PLYP (a popular DH functional) leads to an unphysical divergence at the dissociation limit, due to the vanishing HOMO-LUMO gap 
appeared in the energy denominator of its second-order perturbation energy components. Clearly, hybrid and DH functionals, the most popular schemes for reducing the SIEs and NCIEs of KS-DFAs, 
respectively, can perform poorly for multi-reference systems due to their inaccurate treatment of strong static correlation effects \cite{SciYang,SCE}. 

\begin{figure}
\caption{\label{h2th} Potential energy curves (in total energy) for the ground state of H$_2$, calculated by spin-restricted TAO-LDA (with various $\theta$). The $\theta=0$ case corresponds to spin-restricted KS-LDA.}
\includegraphics[scale=1.0]{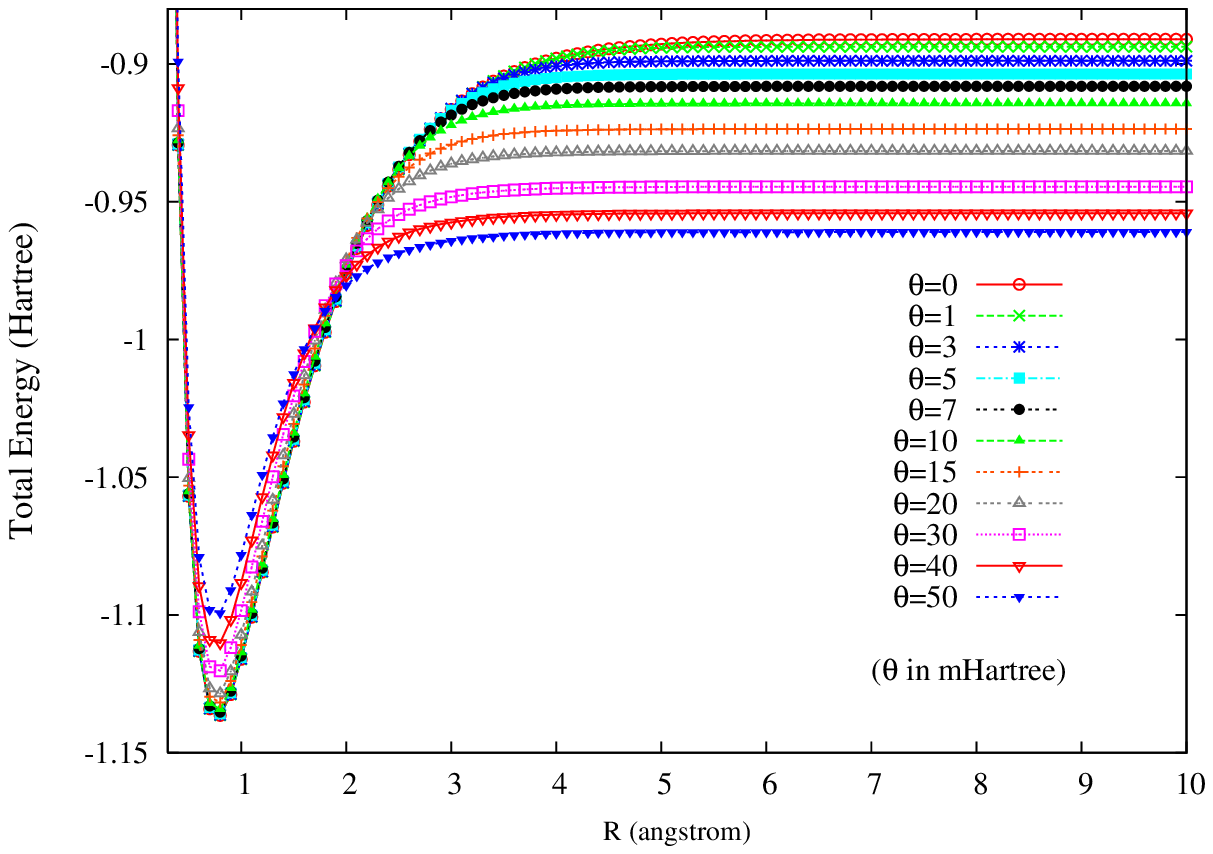}
\end{figure}

\begin{figure}
\caption{\label{h2} Same as Fig.\ \ref{h2th}, but in relative energy. The zeros of energy are set at the respective spin-unrestricted dissociation limits.}
\includegraphics[scale=1.0]{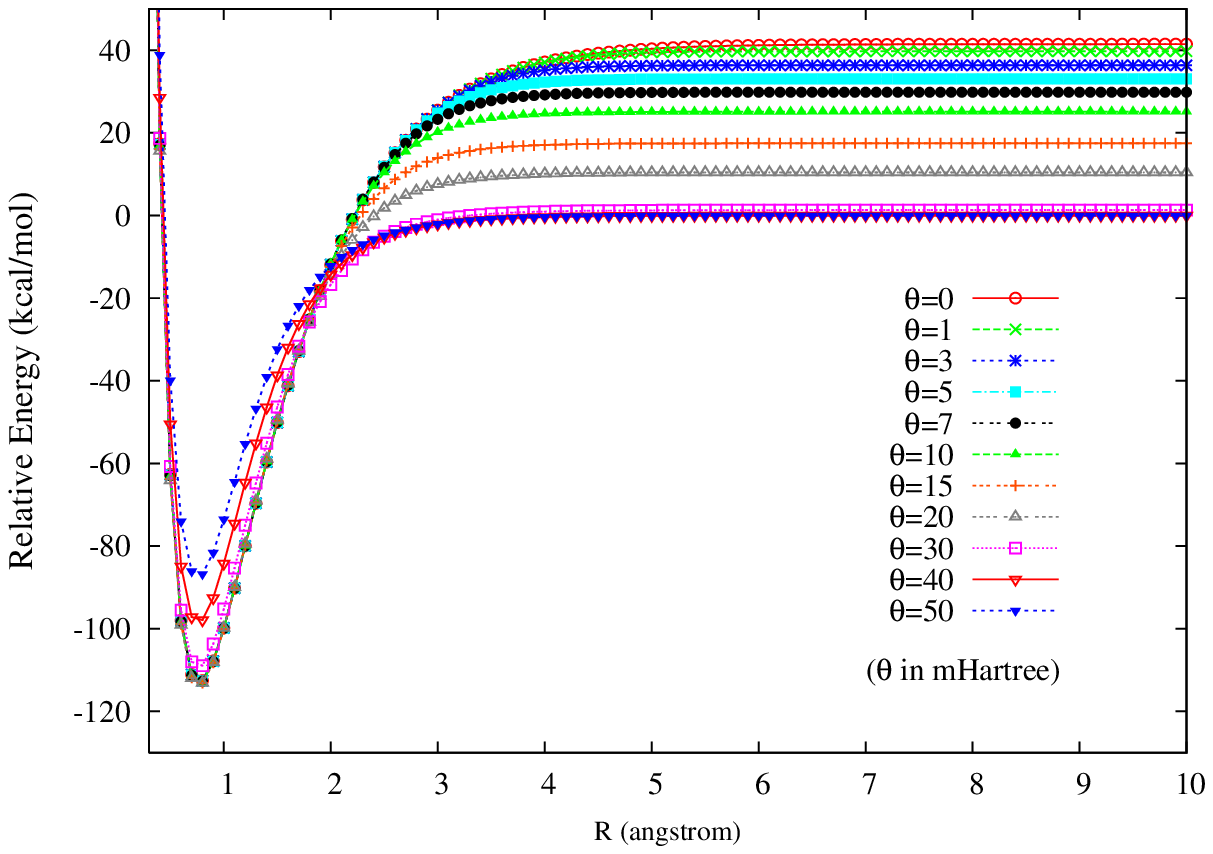}
\end{figure}

To evaluate the performance of the present method, the potential energy curves for the ground state of H$_2$, calculated by spin-restricted TAO-LDA, are shown in Fig.\ \ref{h2th} (in total energy) and 
Fig.\ \ref{h2} (in relative energy). Near the equilibrium geometry, where the multi-reference character is insignificant, the performance of TAO-LDA (with a $\theta$ smaller than 10 mHartree) is very similar to 
that of KS-LDA. At the dissociation limit, where the multi-reference character is pronounced, the SCE of TAO-LDA is shown to be reducible with the increase of $\theta$ value, at essentially no extra computational 
cost! Interestingly, TAO-LDA (with a $\theta$ between 30 and 50 mHartree) performs very well, leading to a vanishingly small SCE. 

\begin{figure}
\caption{\label{h2no} Occupation numbers of the $1\sigma_g$ orbital for the ground state of H$_2$ as a function of the internuclear distance $R$, calculated by spin-restricted TAO-LDA (with various $\theta$). 
The $\theta=0$ case corresponds to spin-restricted KS-LDA. The reference data are the FCI NOONs \cite{H2_NOON}.}
\includegraphics[scale=1.0]{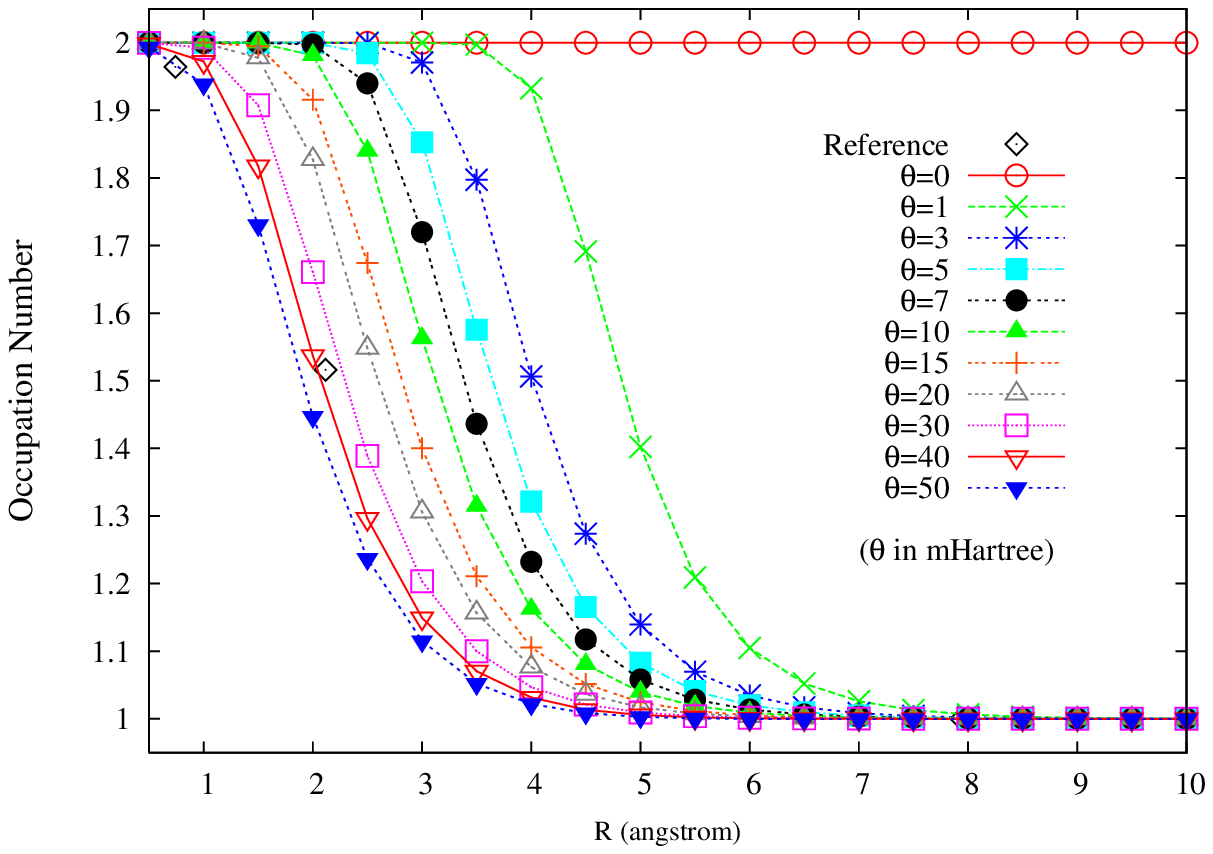}
\end{figure}

To see how this is related to the ensemble representation (via fractional orbital occupations) of the ground-state density, the occupation numbers of the $1\sigma_g$ orbital (HOMO) for the ground state of H$_2$ 
as a function of the internuclear distance $R$, calculated by spin-restricted TAO-LDA, are presented in Fig.\ \ref{h2no}, where the reference data are the FCI NOONs \cite{H2_NOON}. 
At the equilibrium geometry ($R$ = 0.741 {\AA}), the FCI NOON is 1.9643, indicating the absence of strong static correlation effects (with doubly occupied $1\sigma_g$ orbital). However, 
the FCI NOON is 1.5162 at $R$ = 2.117 {\AA}, and 1.0000 at $R$ = 7.938 {\AA}, indicating the presence of strong static correlation effects. The $1\sigma_g$ orbital occupation numbers of spin-restricted TAO-LDA 
(with a nonvanishing $\theta$) are very close to 2.0 (doubly occupied) near the equilibrium geometry, and gradually reduced to 1.0 (singly occupied) at the dissociation limit. The larger the $\theta$ value is, 
the faster the corresponding $1\sigma_g$ orbital occupation number approaches 1.0 at the larger $R$. The $1\sigma_g$ orbital occupation numbers of spin-restricted TAO-LDA (with a $\theta$ between 30 and 50 mHartree) 
are shown to match well with the FCI NOONs, which is closely related to the vanishingly small SCE of TAO-LDA (with the same $\theta$). 

\begin{figure}
\caption{\label{h2entr} Potential energy curves (in relative energy) for the ground state of H$_2$, calculated by the spin-restricted (with and without the entropy contributions) and spin-unrestricted 
TAO-LDA ($\theta$ = 40 mHartree), where the zeros of energy are set at the spin-unrestricted dissociation limit. The entropy contributions (in total energy) as a function of the internuclear distance $R$, 
calculated by spin-restricted TAO-LDA ($\theta$ = 40 mHartree), are also shown.}
\includegraphics[scale=1.0]{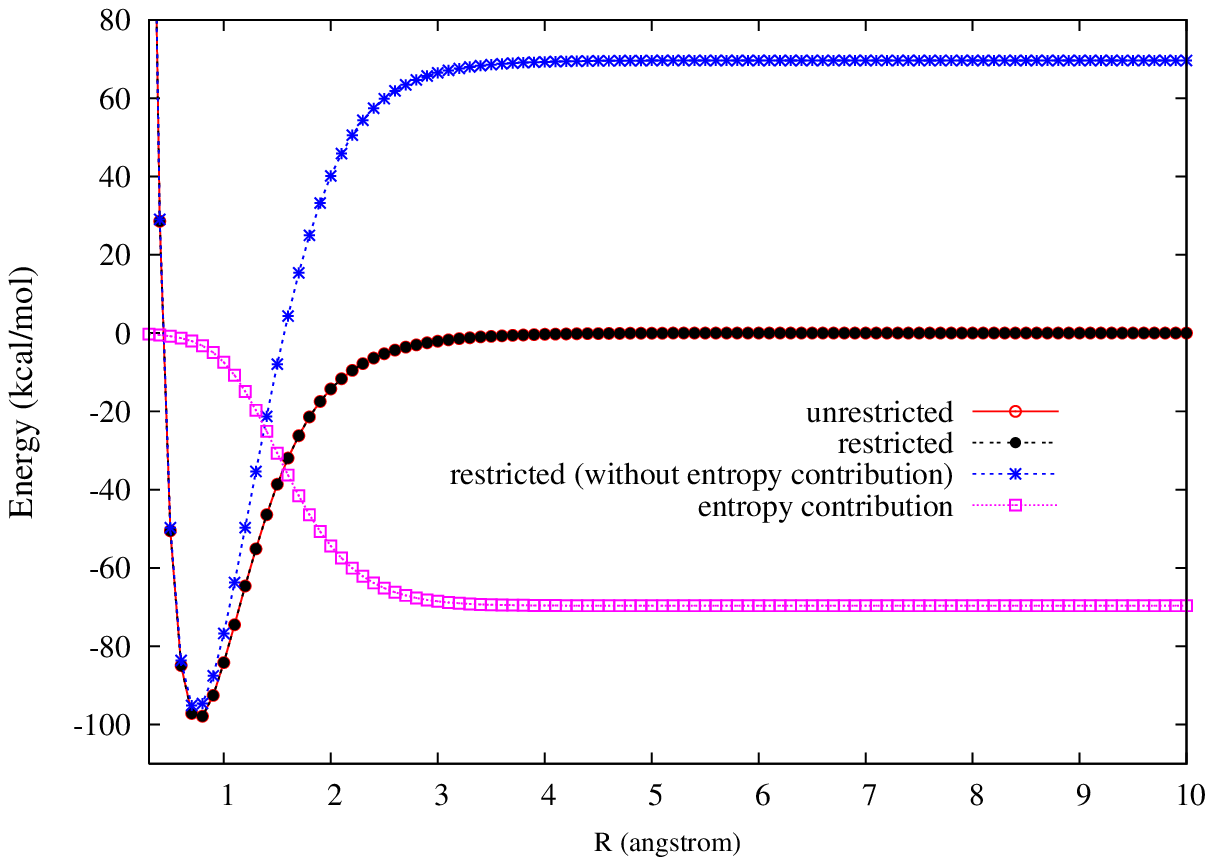}
\end{figure}

\begin{figure}
\caption{\label{h2entr2} Same as Fig.\ \ref{h2entr}, but for $\theta$ = 7 mHartree.} 
\includegraphics[scale=1.0]{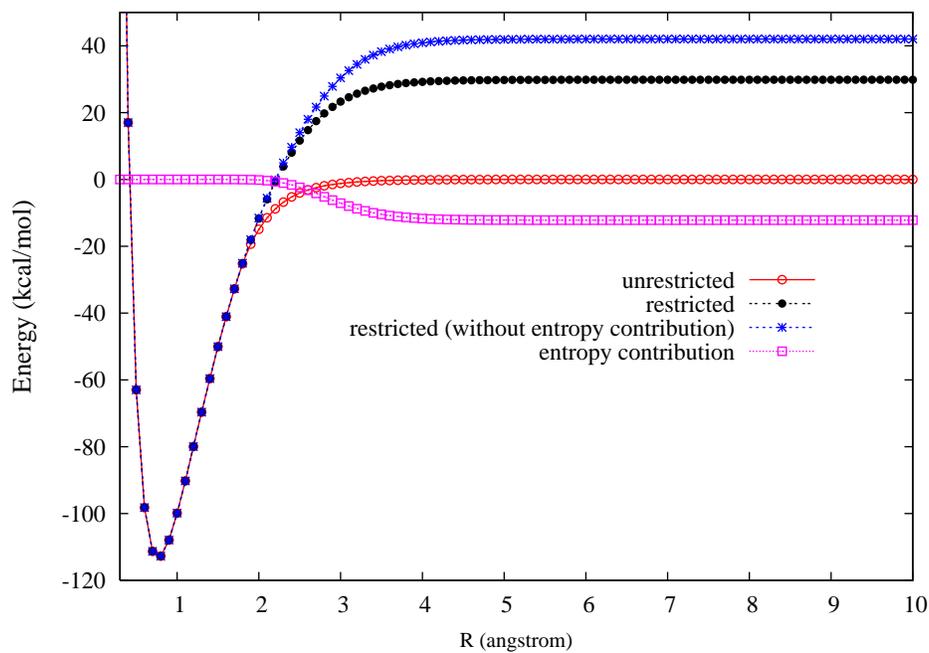}
\end{figure}

To examine the entropy contributions (in total energy) as a function of the internuclear distance $R$, calculated by spin-restricted TAO-LDA ($\theta$ = 40 mHartree), Fig.\ \ref{h2entr} shows the potential energy 
curves (in relative energy) for the ground state of H$_2$, calculated by the spin-restricted (with and without the entropy contributions) and spin-unrestricted TAO-LDA ($\theta$ = 40 mHartree), where the zeros of energy 
are set at the spin-unrestricted dissociation limit. As can be seen, the entropy contributions are insignificant near the equilibrium geometry and pronounced at the 
larger $R$ (approaching a negative constant at the dissociation limit), properly simulating the strong static correlation effects to make the spin-restricted potential energy curve the same as the 
spin-unrestricted one (as it should be). By contrast, the entropy contributions as a function of the internuclear distance $R$, calculated by spin-restricted TAO-LDA ($\theta$ = 7 mHartree), are still insufficient to simulate 
the strong static correlation effects (see Fig.\ \ref{h2entr2}), as the corresponding $1\sigma_g$ orbital occupation numbers do not match well with the FCI NOONs (see Fig.\ \ref{h2no}).

\begin{figure}
\caption{\label{n2th} Potential energy curves (in total energy) for the ground state of N$_2$, calculated by spin-restricted TAO-LDA (with various $\theta$). The $\theta=0$ case corresponds to spin-restricted KS-LDA.}
\includegraphics[scale=1.0]{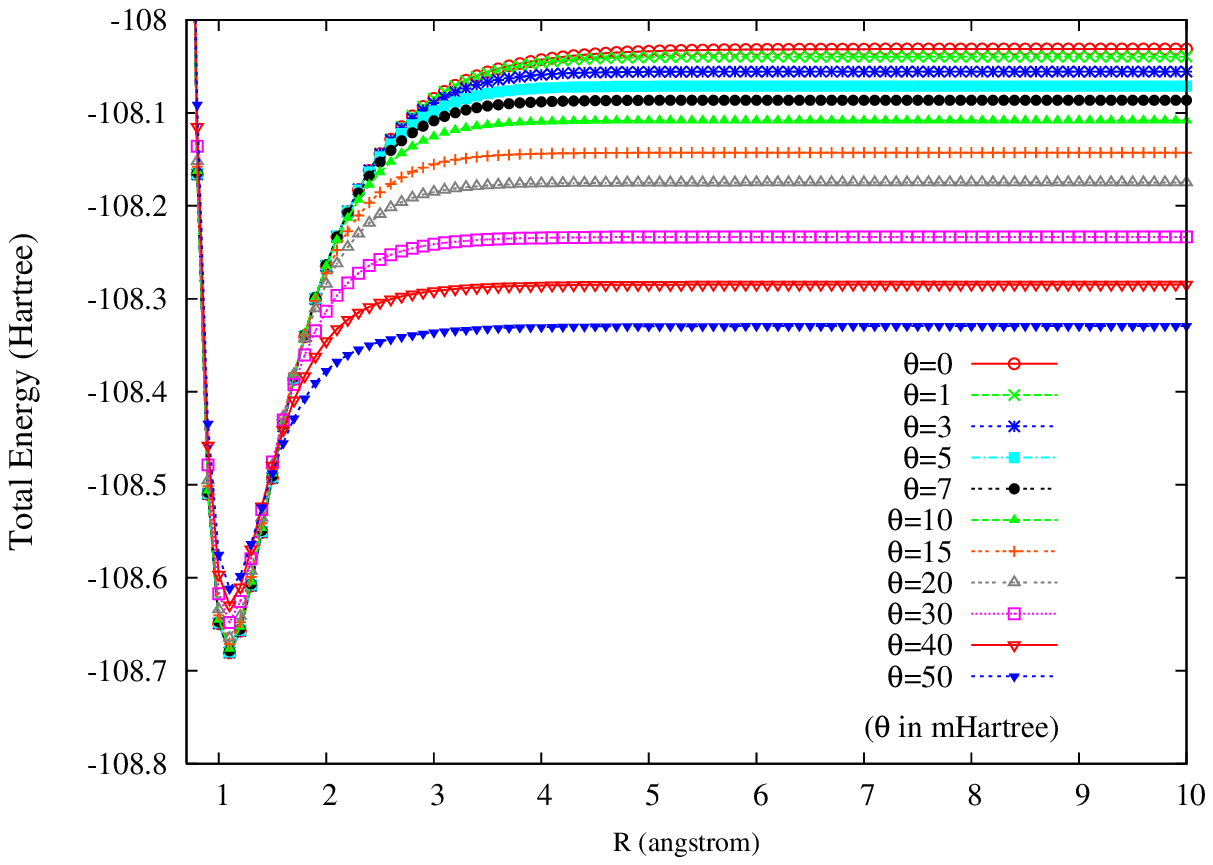}
\end{figure}

\begin{figure}
\caption{\label{n2} Same as Fig.\ \ref{n2th}, but in relative energy. The zeros of energy are set at the respective spin-unrestricted dissociation limits.}
\includegraphics[scale=1.0]{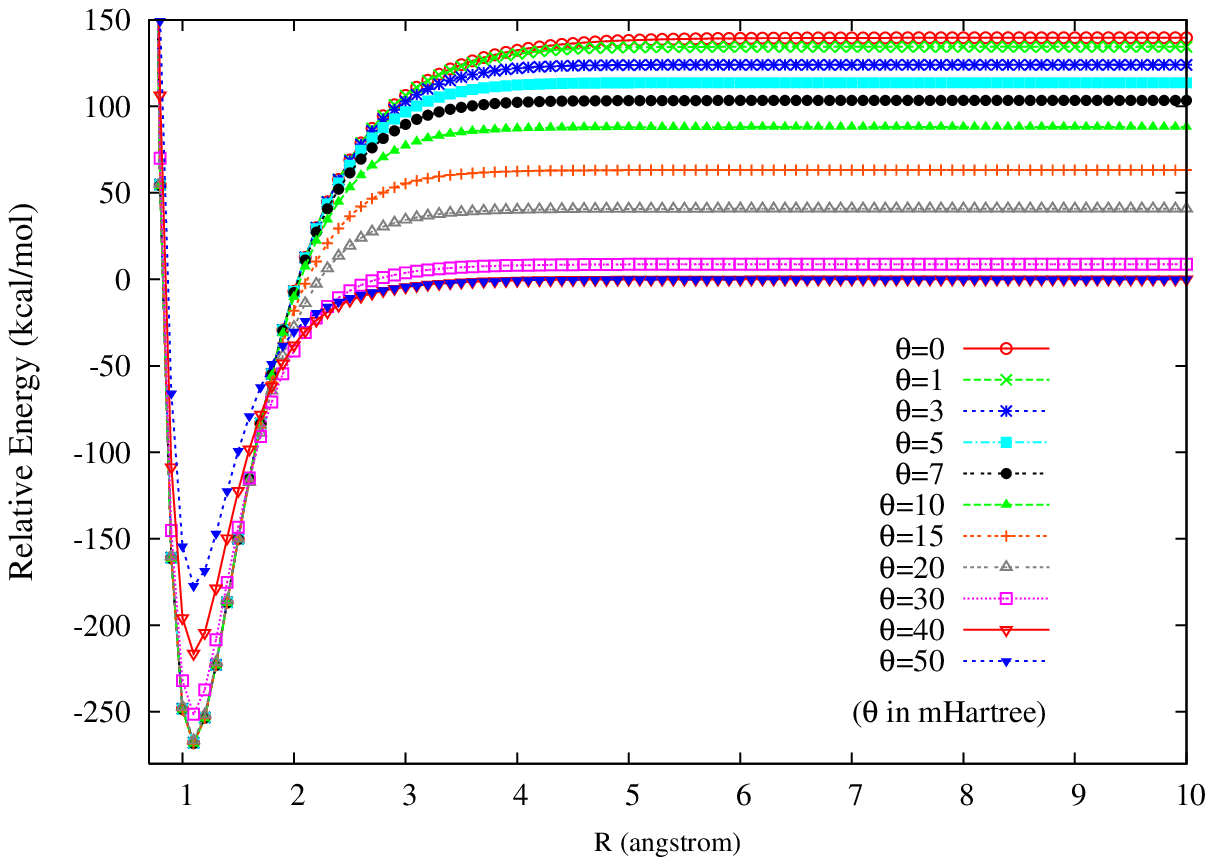}
\end{figure}

\begin{figure}
\caption{\label{n2nog} Occupation numbers of the $3\sigma_g$ orbital for the ground state of N$_2$ as a function of the internuclear distance $R$, calculated by spin-restricted TAO-LDA (with various $\theta$). 
The $\theta=0$ case corresponds to spin-restricted KS-LDA. The reference data are the MRCI NOONs \cite{N2_NOON}.}
\includegraphics[scale=1.0]{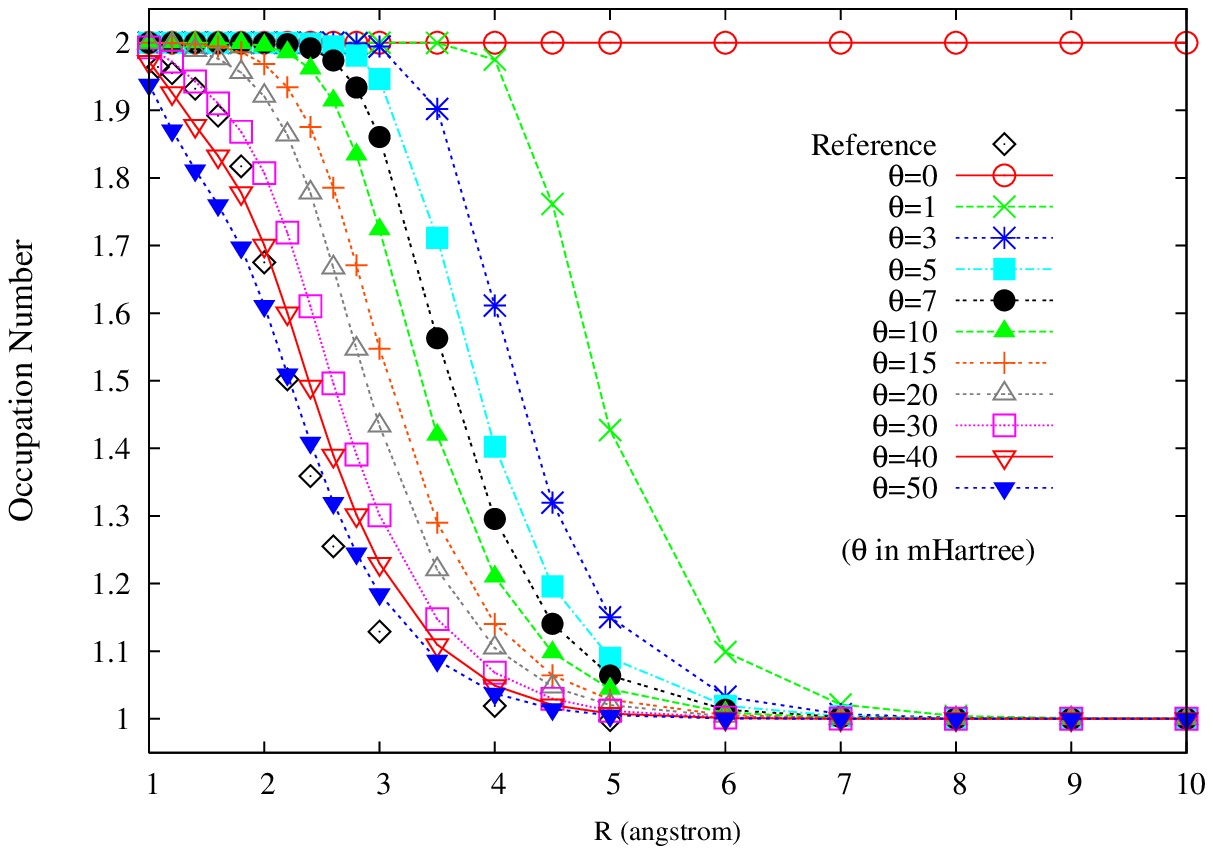}
\end{figure}

\begin{figure}
\caption{\label{n2nopi} Same as Fig.\ \ref{n2nog}, but for the 1$\pi_{ux}$ orbital.}
\includegraphics[scale=1.0]{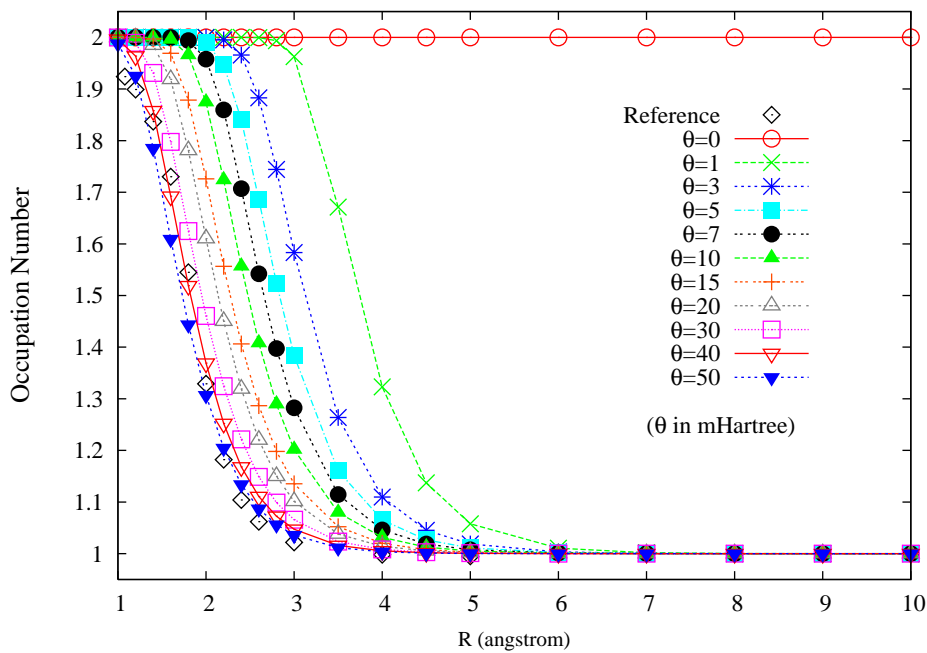}
\end{figure}

Similar results are also found for N$_2$ dissociation, a triple-bond breaking system. The potential energy curves for the ground state of N$_2$, calculated by spin-restricted TAO-LDA, are shown in 
Fig.\ \ref{n2th} (in total energy) and Fig.\ \ref{n2} (in relative energy). As can be seen, spin-restricted TAO-LDA (with a $\theta$ between 30 and 50 mHartree) can dissociate N$_2$ properly 
(yielding a vanishingly small SCE) to the respective spin-unrestricted dissociation limits, which is closely related to that the occupation numbers of the 3$\sigma_g$ (in Fig.\ \ref{n2nog}) 
and 1$\pi_{ux}$ (in Fig.\ \ref{n2nopi}) orbitals for the ground state of N$_2$ as functions of the internuclear distance $R$, calculated by spin-restricted TAO-LDA (with the same $\theta$), match reasonably well with 
the corresponding MRCI NOONs (the reference data) \cite{N2_NOON}. 

\begin{figure}
\caption{\label{n2entr} Potential energy curves (in relative energy) for the ground state of N$_2$, calculated by the spin-restricted (with and without the entropy contributions) and spin-unrestricted 
TAO-LDA ($\theta$ = 40 mHartree), where the zeros of energy are set at the spin-unrestricted dissociation limit. The entropy contributions (in total energy) as a function of the internuclear distance $R$, calculated by 
spin-restricted TAO-LDA ($\theta$ = 40 mHartree), are also shown.}
\includegraphics[scale=1.0]{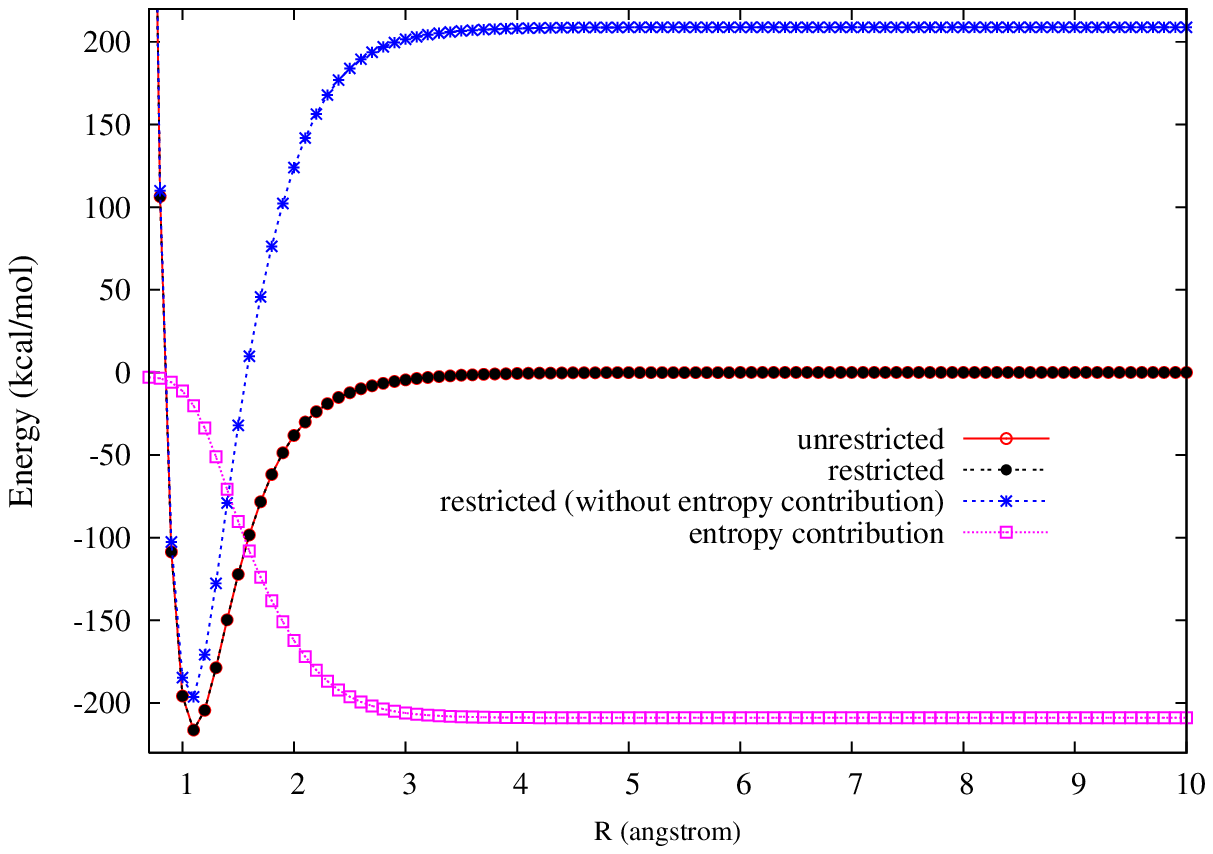}
\end{figure}

To examine the entropy contributions (in total energy) as a function of the internuclear distance $R$, calculated by spin-restricted TAO-LDA ($\theta$ = 40 mHartree), Fig.\ \ref{n2entr} shows the potential energy 
curves (in relative energy) for the ground state of N$_2$, calculated by the spin-restricted (with and without the entropy contributions) and spin-unrestricted TAO-LDA ($\theta$ = 40 mHartree), where the zeros of energy 
are set at the spin-unrestricted dissociation limit. As shown, the entropy contributions are essentially responsible for simulating the strong static correlation effects to make the spin-restricted potential energy curve 
the same as the spin-unrestricted one (as it should be). For spin-restricted TAO-LDA ($\theta$ = 40 mHartree), the entropy contribution (-208.90 kcal/mol) at the dissociation limit of N$_2$, is considerably larger 
(about three times larger) than that (-69.64 kcal/mol) at the dissociation limit of H$_2$, as the number of unpaired electrons (or singly occupied orbitals) for N$_2$ dissociation is more than (three times more) that 
for H$_2$ dissociation. 

To sum up, when the orbital occupation numbers $\{f_i\}$ of TAO-LDA are close to the exact NOONs $\{n_i\}$, the strong static correlation effects are shown to be properly simulated by the entropy contribution 
of TAO-LDA. As this feature is independent of the number of unpaired electrons in a system, TAO-LDA seems to be promising for the study of large polyradical systems, such as linear acenes (as will be shown later).

\subsubsection{Twisted Ethylene}

\begin{figure}
\caption{\label{c2h4} Torsion potential energy curves (in relative energy) for the ground state of twisted ethylene as a function of the HCCH torsion angle, calculated by spin-restricted TAO-LDA (with various $\theta$). 
The zeros of energy are set at the respective minimum energies. The $\theta=0$ case corresponds to KS-LDA.}
\includegraphics[scale=1.0]{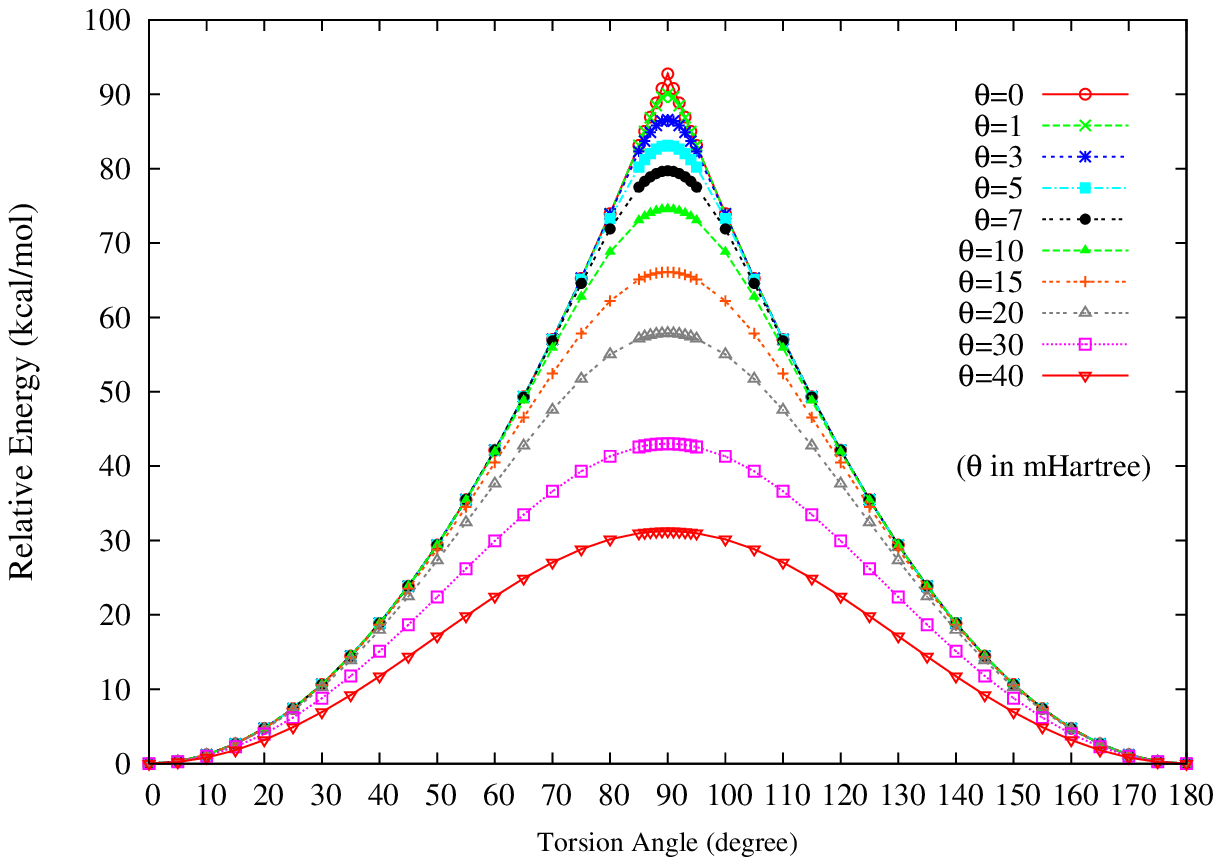}
\end{figure}

The $\pi$ (1b$_2$) and $\pi^*$ (2b$_2$) orbitals in ethylene (C$_2$H$_4$) should be degenerate when the HCCH torsion angle is 90$^{\circ}$. Spin-restricted single-reference methods cannot handle 
such a degeneracy properly and show an unphysical cusp in the torsion potential near 90$^{\circ}$. In the calculations, we use the experimental geometry of 
C$_2$H$_4$ ($R_{\text{CC}}$ = 1.339 {\AA}, $R_{\text{CH}}$ = 1.086 {\AA}, $\angle_{\text{HCH}}$ = 117.6$^{\circ}$) \cite{C2H4_Geom}. Fig.\ \ref{c2h4} shows the torsion potential energy curves (in relative energy) 
for the ground state of twisted ethylene as a function of the HCCH torsion angle, calculated by spin-restricted TAO-LDA, where the zeros of energy are set at the respective minimum energies. Spin-restricted 
TAO-LDA (with a $\theta$ larger than 5 mHartree) is shown to be able to remove the unphysical cusp, though spin-restricted TAO-LDA (with a $\theta$ larger than 20 mHartree) is shown to yield a torsion barrier 
which is far too low. 

\begin{figure}
\caption{\label{c2h4noon} Occupation numbers of the $\pi$ (1b$_2$) orbital for the ground state of twisted ethylene as a function of the HCCH torsion angle, calculated by spin-restricted TAO-LDA (with various $\theta$). 
The $\theta=0$ case corresponds to spin-restricted KS-LDA. The reference data are the half-projected NOONs of CASSCF method (HPNO-CAS) \cite{C2H4_NOON}.}
\includegraphics[scale=1.0]{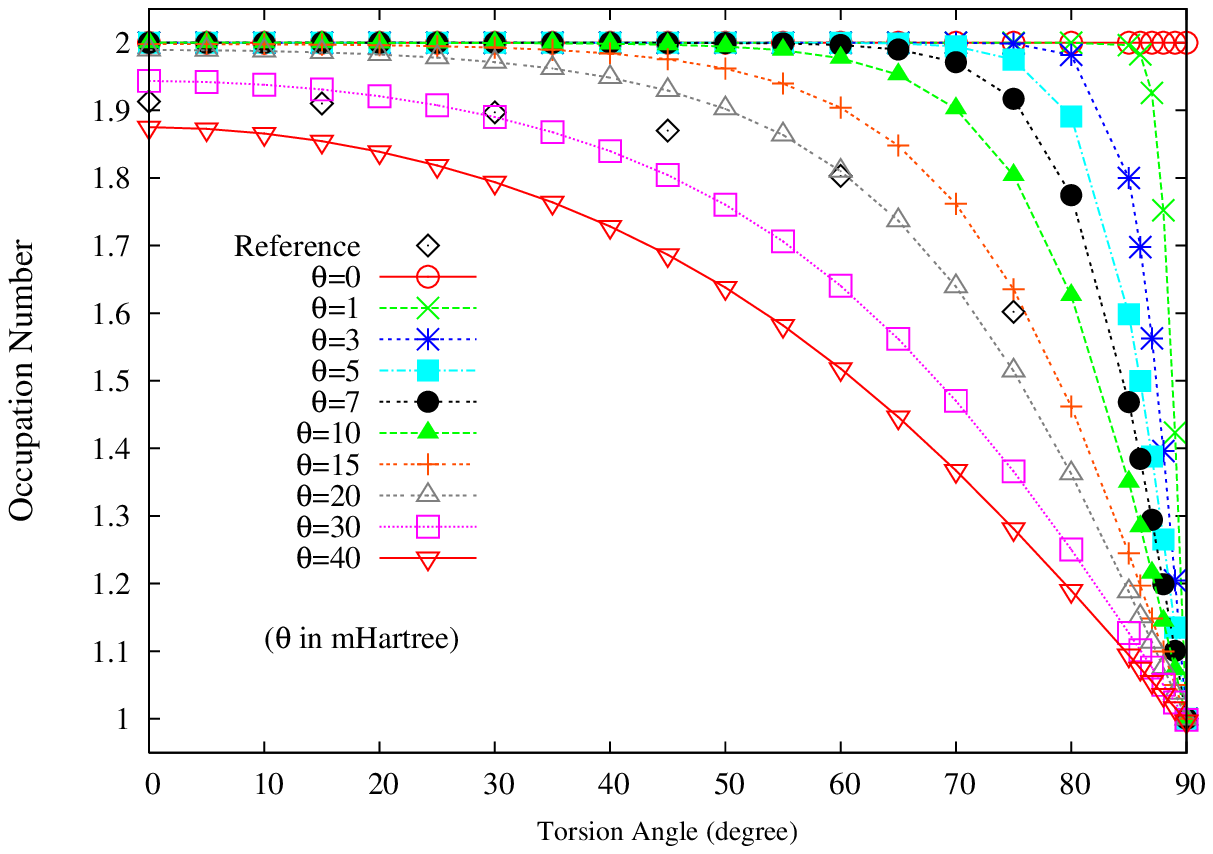}
\end{figure}

Fig.\ \ref{c2h4noon} shows the occupation numbers of the $\pi$ (1b$_2$) orbital for the ground state of twisted ethylene as a function of the HCCH torsion angle, calculated by spin-restricted TAO-LDA, where 
the reference data are the half-projected NOONs of CASSCF method (HPNO-CAS) \cite{C2H4_NOON}. As can be seen, the $\pi$ (1b$_2$) orbital occupation numbers of spin-restricted 
TAO-LDA (with a $\theta$ between 10 and 20 mHartree), match reasonably well with the accurate NOONs, which is related to the accurate torsion potential energy curve, calculated by spin-restricted 
TAO-LDA (with the same $\theta$). 

\begin{figure}
\caption{\label{c2h4entr} Torsion potential energy curves (in relative energy) for the ground state of twisted ethylene as a function of the HCCH torsion angle, calculated by the spin-restricted 
(with and without the entropy contributions) and spin-unrestricted TAO-LDA ($\theta$ = 15 mHartree), where the zeros of energy are set at the respective minimum energies. 
The entropy contributions (in total energy) as a function of the HCCH torsion angle, calculated by spin-restricted TAO-LDA ($\theta$ = 15 mHartree), are also shown.}
\includegraphics[scale=1.0]{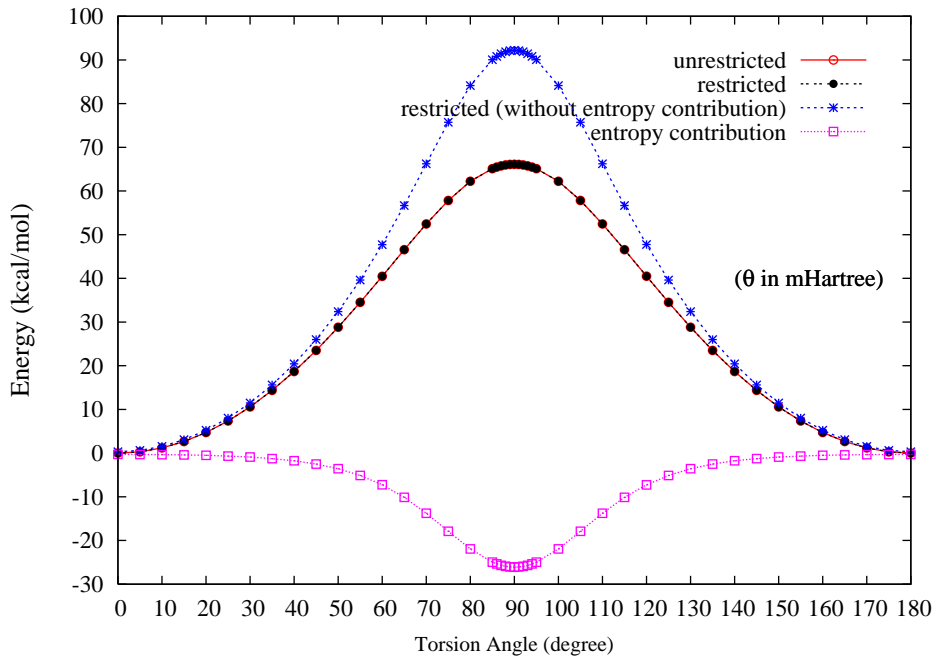}
\end{figure}

To examine the entropy contributions (in total energy) as a function of the HCCH torsion angle, calculated by spin-restricted TAO-LDA ($\theta$ = 15 mHartree), Fig.\ \ref{c2h4entr} shows the torsion potential energy 
curves (in relative energy) for the ground state of twisted ethylene as a function of the HCCH torsion angle, calculated by the spin-restricted (with and without the entropy contributions) and spin-unrestricted 
TAO-LDA ($\theta$ = 15 mHartree), where the zeros of energy are set at the respective minimum energies. As shown, the entropy contributions are responsible for simulating the strong static correlation effects to make 
the spin-restricted torsion potential energy curve the same as the spin-unrestricted one (as it should be).

\subsubsection{Singlet-Triplet Energy Gaps of Linear Acenes}

\begin{figure}
\caption{\label{pentacene} Pentacene, consisting of 5 linearly fuzed benzene rings, is designated as 5-acene.\ }
\includegraphics[scale=1.0]{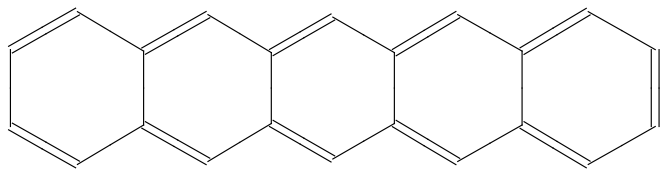}
\end{figure}

Linear $n$-acenes (C$_{4n+2}$H$_{2n+4}$), consisting of $n$ linearly fused benzene rings (see Fig.\ (\ref{pentacene})), have attracted great interest from many experimental and theoretical researchers 
due to their fascinating electronic properties and technological potential \cite{2-acene,3-acene,4-acene,5-acene,aceneHouk,aceneRaghu,aceneCarter,aceneSantos,aceneChan,aceneJiang,aceneIshida,
aceneLiu,aceneHajgato,aceneYang,aceneHajgato2,aceneMazziotti}. The experimental singlet-triplet energy gaps (ST gaps) of $n$-acenes are only available up to pentacene \cite{2-acene,3-acene,4-acene,
5-acene}, due to the increasing reactivity of the larger acenes. Recently, the calculated ST gaps have been in serious 
debate \cite{aceneHouk,aceneRaghu,aceneCarter,aceneSantos,aceneChan,aceneJiang,aceneIshida,aceneLiu,aceneHajgato,aceneYang,aceneHajgato2}. Typically, delocalized $\pi$-orbital systems, 
such as $n$-acenes, require high-level {\it ab initio} methods, such as the DMRG algorithm \cite{aceneChan}, to capture the essential strong static correlation effects. Based on the recent work of Chan and 
co-workers \cite{aceneChan}, the DMRG ST gaps as a function of the acene length have been shown to decrease monotonically with increasing chain length. Based on a good fit to the DMRG ST gaps of the smaller 
$n$-acenes (up to 12-acene), an exponential fitting function of the form $a + b \ {\text e}^{-c \ n}$ was adopted for extrapolation of the ST gaps to the infinite chain limit \cite{aceneChan}, yielding a finite 
ST gap (3.33 kcal/mol for the cc-pVDZ basis set, and 8.69 kcal/mol for the STO-3G basis set) for polyacene (triplet above singlet). 
However, the extrapolated results have been shown subject to details of the fit \cite{aceneHajgato}. More importantly, it is unclear whether the ST gaps of the larger $n$-acenes (eg.\ $n \ge 20$) still 
decrease exponentially (as those of the smaller $n$-acenes) with increasing chain length, which may significantly affect the extrapolated ST gap. Calculations on the larger acenes are necessary to address this, 
which are, however, prohibitively expensive for the DMRG algorithm \cite{aceneChan} and other high-level {\it ab initio} methods \cite{aceneHajgato,aceneHajgato2}. 

\begin{figure}
\caption{\label{f3a} Singlet-triplet energy gap as a function of the acene length, calculated by spin-unrestricted KS-DFT (with LDA, BLYP, and B3LYP functionals), using 
the 6-31G* basis set. The experimental data are taken from Refs.\ \cite{2-acene,3-acene,4-acene,5-acene}, and the DMRG data are taken from Ref.\ \cite{aceneChan}.}
\includegraphics[scale=1.0]{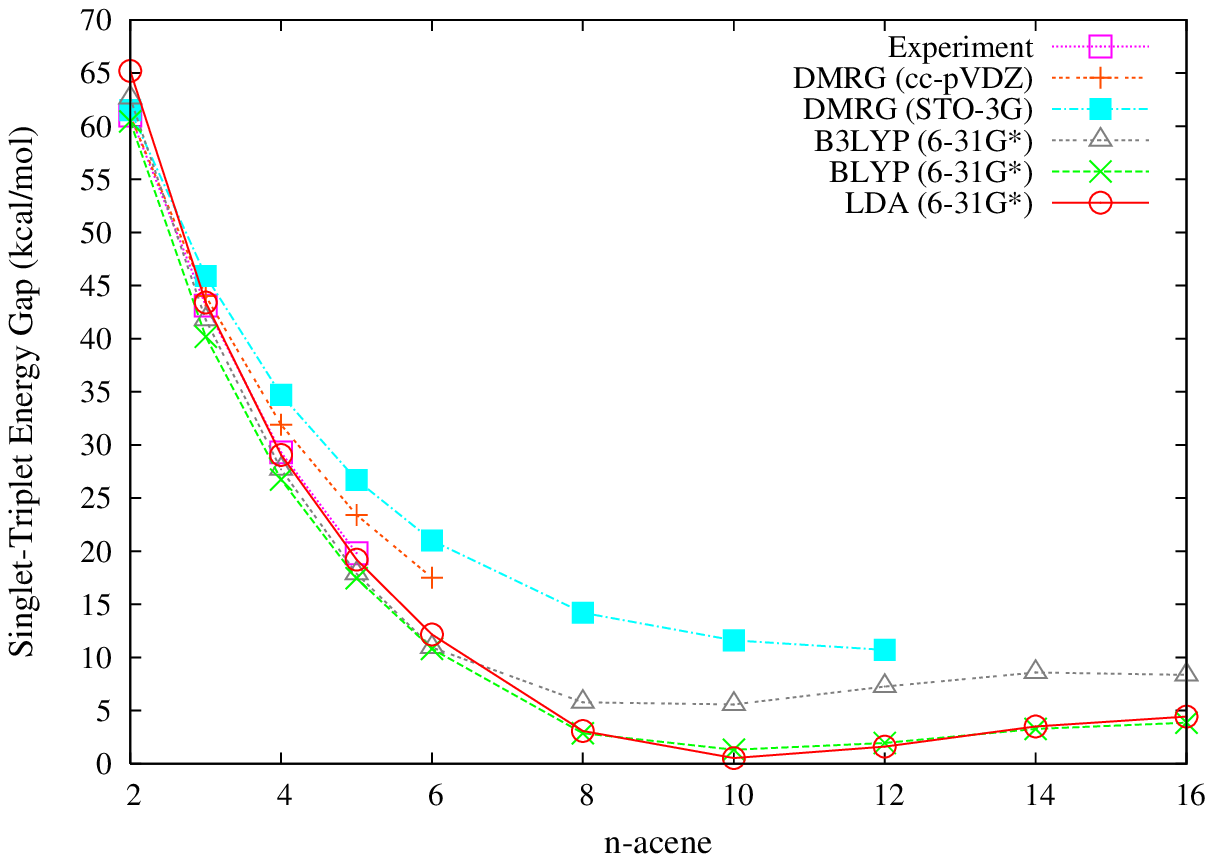}
\end{figure}

On the other hand, KS-DFT is computationally efficient, but unable to handle such strong static correlation effects properly. To show this, spin-unrestricted KS-DFT 
calculations (with LDA \cite{LDAX,LDAC}, BLYP \cite{B88,LYP}, and B3LYP \cite{hybrid,B3LYP} functionals) 
are performed, using the 6-31G* basis set (up to 16-acene), for the lowest singlet and triplet energies on the respective geometries that were fully optimized at the same level. The ST gap of $n$-acene is 
calculated as $(E_{\text{T}} - E_{\text{S}})$, the energy difference between the lowest triplet (T) and singlet (S) states of $n$-acene. As shown in Fig.\ (\ref{f3a}), in contrast to the DMRG results \cite{aceneChan}, 
the ST gaps calculated by spin-unrestricted KS-DFT, are shown to unexpectedly increase beyond 10-acene, due to unphysical symmetry-breaking effects \cite{supp}. 

\begin{figure}
\caption{\label{f3} Singlet-triplet energy gap as a function of the acene length, calculated by spin-unrestricted TAO-LDA (with various $\theta$), using the 6-31G* basis set. 
The $\theta=0$ case corresponds to spin-unrestricted KS-LDA. The experimental data are taken from Refs. \cite{2-acene,3-acene,4-acene,5-acene}.}
\includegraphics[scale=1.0]{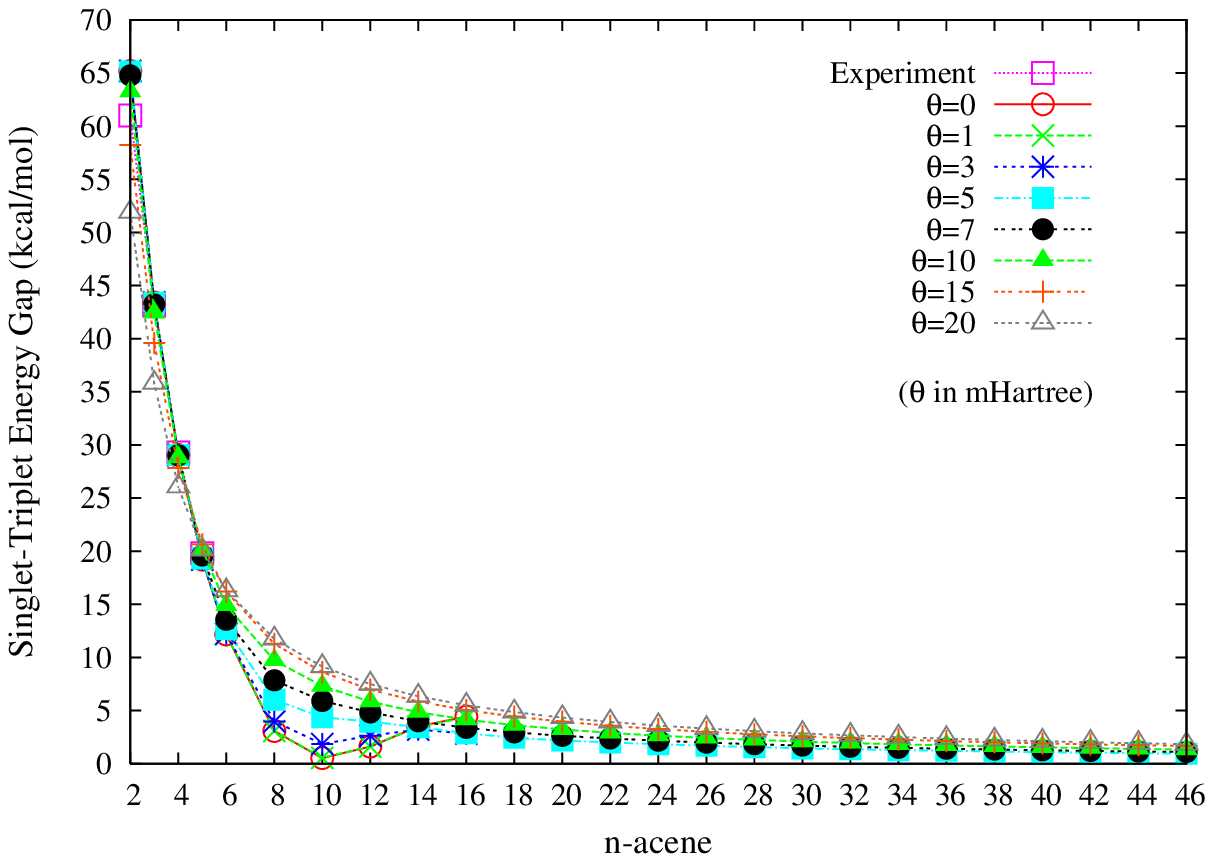}
\end{figure}

To assess the performance of the present method, spin-unrestricted TAO-LDA calculations are performed, using both the 6-31G* (up to 46-acene) and 6-31G (up to 74-acene) basis sets, for the lowest singlet and triplet 
energies on the respective geometries that were fully optimized at the same level \cite{supp}. In Fig.\ (\ref{f3}), the calculated ST gaps as a function of the acene length (using the 6-31G* basis set) are plotted. 
In contrast to the spin-unrestricted KS-LDA results, the ST gaps calculated by spin-unrestricted TAO-LDA (with a $\theta$ larger than 5 mHartree), are shown to decrease monotonically with the increase of chain length. 
The ST gaps calculated by spin-unrestricted TAO-LDA (with a $\theta$ between 5 and 10 mHartree), are in good agreement with the existing experimental and high-level {\it ab initio} 
data \cite{aceneChan,aceneHajgato,aceneHajgato2}. Due to the symmetry constraint, the spin-restricted and spin-unrestricted energies for the lowest singlet state of $n$-acene, calculated by the exact theory, should be the same. 
To examine this property, spin-restricted TAO-LDA calculations (using the 6-31G* basis set) are also performed for the lowest singlet energies on the respective geometries that were fully optimized at the same level. 
The spin-unrestricted and spin-restricted TAO-LDA (with a $\theta$ larger than 5 mHartree) calculations are found to essentially yield the same energy value for the lowest singlet state of $n$-acene 
(i.e.\ no unphysical symmetry-breaking effects). 

\begin{figure}
\caption{\label{f4} Singlet-triplet energy gap as a function of the acene length, calculated by spin-unrestricted TAO-LDA ($\theta$ = 7 mHartree), using both the 6-31G* and 6-31G basis sets.}
\includegraphics[scale=1.0]{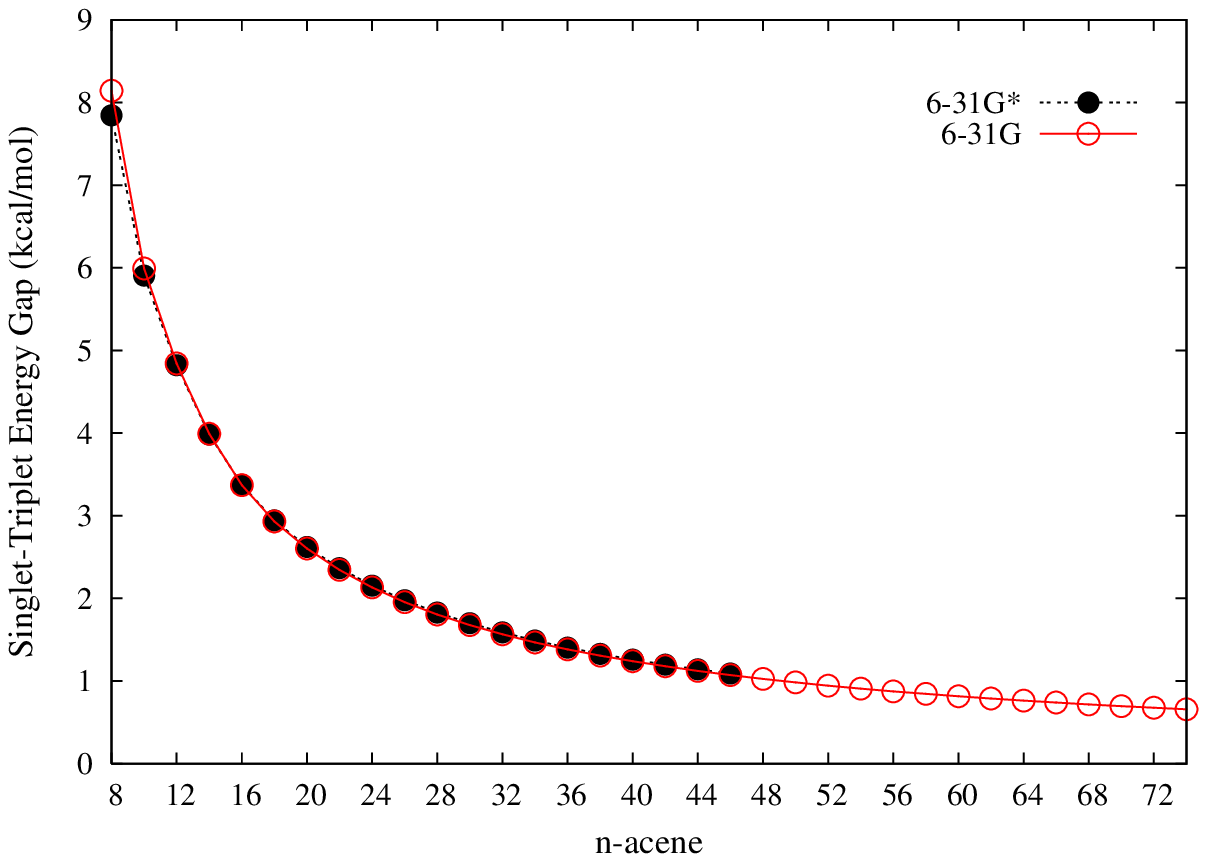}
\end{figure}

Fig.\ (\ref{f4}) shows the ST gaps of $n$-acenes ($n \ge 8$) as a function of the acene length, calculated by spin-unrestricted TAO-LDA ($\theta$ = 7 mHartree), using both the 6-31G* and 6-31G basis sets \cite{supp}. 
The effects of basis sets on the calculated ST gaps are shown to be insignificant for the larger acenes. The ground state of $n$-acene is found to remain a singlet as chain length is increased. 
At the level of spin-unrestricted TAO-LDA ($\theta$ = 7 mHartree)/6-31G, the ST gap of the largest acene studied here (74-acene) is 0.66 kcal/mol. In view of the slow convergence of the ST gaps with the increase of chain length, 
the ST gaps of the larger $n$-acenes ($n \ge 20$) are found to fit extremely well to a power-law function of the form $a + b \ n^{-c}$, rather than the popular exponential function \cite{aceneChan,aceneHajgato,aceneHajgato2}. 
As shown in Table \ref{table:acenes3a}, nonlinear least-squares fitting of 3 different data sets ($20$- to $74$-acene, $30$- to $74$-acene, and $40$- to $74$-acene) of the ST gaps calculated by spin-unrestricted 
TAO-LDA ($\theta$ = 7 mHartree)/6-31G, by means of the above power-law fitting function, gives estimates of 0.08, 0.04, and 0.03 (kcal/mol), respectively, for the ST gaps of $n$-acenes in the polymer limit ($n \to \infty$). 
As the extrapolated ST gaps are rather insensitive to the choices of the fitting data sets, the ST gaps selected in the fitting data sets should have approached the asymptotic (large-$n$) behavior, decreasing as slowly as 
about $n^{-1}$ with increasing chain length. In view of the minor dependence of the extrapolated results with the choices of the fitting data sets, we can only conclude that in the polymer limit, the lowest singlet and triplet states 
should be degenerate within 0.1 kcal/mol (triplet above singlet), which supports the absence of Peierls distortions \cite{Peierls} in this limit, and the closure of the fundamental gap \cite{aceneRaghu,aceneHajgato}.  

\begin{figure}
\caption{\label{acene_ho} HOMO occupation numbers for the lowest singlet states of $n$-acenes as a function of the acene length, calculated by spin-restricted TAO-LDA (with various $\theta$)/6-31G*. 
The $\theta=0$ case corresponds to spin-restricted KS-LDA. Reference data are the NOONs of the active-space variational 2-RDM method \cite{aceneMazziotti}.}
\includegraphics[scale=1.0]{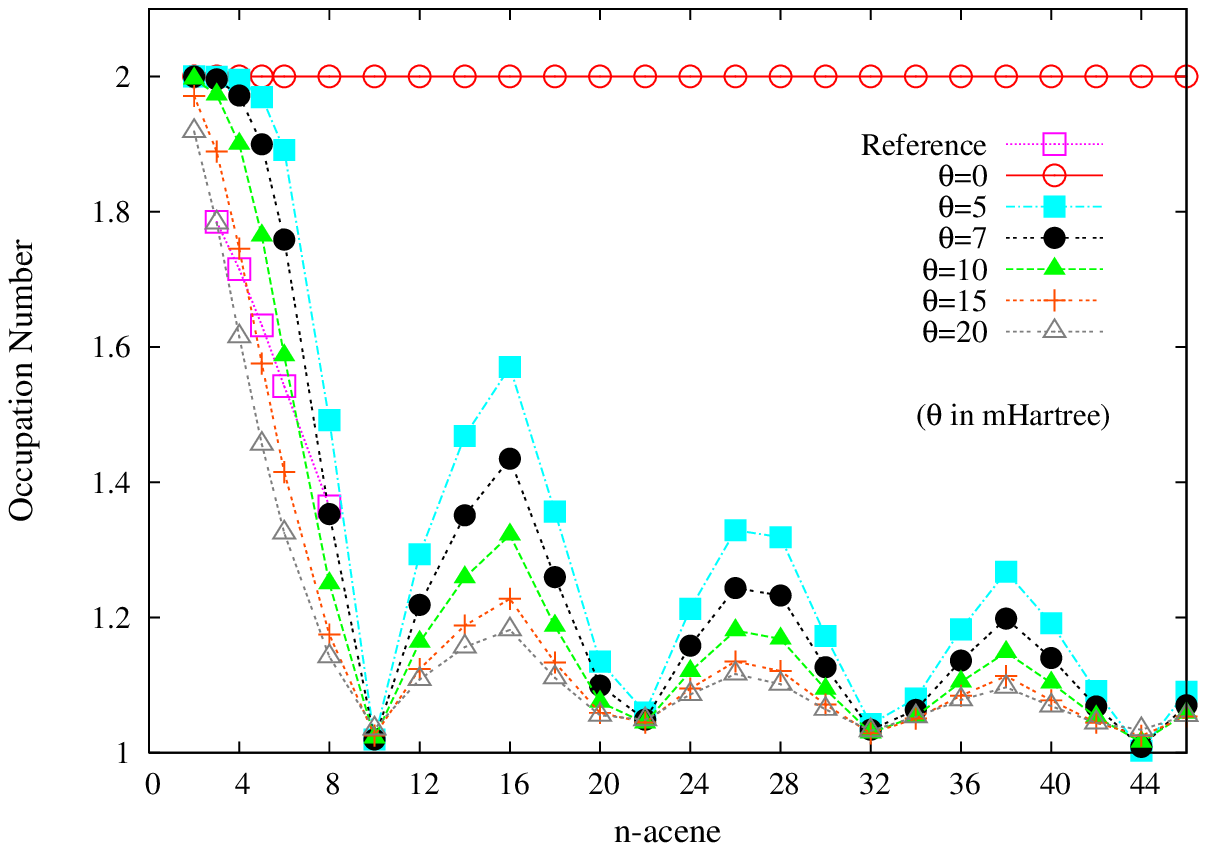}
\end{figure}

The orbital occupation numbers of TAO-LDA provide information useful in assessing the possible polyradical character of $n$-acenes. Fig.\ (\ref{acene_ho}) shows the HOMO occupation numbers for the lowest singlet states 
of $n$-acenes as a function of the acene length, calculated by spin-restricted TAO-LDA/6-31G*, where the reference data are the NOONs of the active-space variational 2-RDM method \cite{aceneMazziotti}. 
Here, HOMO is the ${(N/2)}^{th}$ orbital, and LUMO is the ${(N/2 + 1)}^{th}$ orbital, where $N$ is the number of electrons in $n$-acene. As can be seen, the HOMO occupation numbers of spin-restricted 
TAO-LDA (with a $\theta$ between 5 and 15 mHartree), match reasonably well with the NOONs, which may suggest that the ST gaps calculated by TAO-LDA (with a $\theta$ between 5 and 15 mHartree) should be 
reliably accurate (due to the appropriate treatment of strong static correlation effects via the entropy contribution), providing that these agreements are extendible for the larger acenes. 

\begin{figure}
\caption{\label{acene_noon} Active orbital occupation numbers (HOMO-6, ..., HOMO-1, HOMO, LUMO, LUMO+1, ..., and LUMO+6) for the lowest singlet states of $n$-acenes as a function of the acene length, 
calculated by spin-restricted TAO-LDA ($\theta$ = 7 mHartree)/6-31G*.}
\includegraphics[scale=1.0]{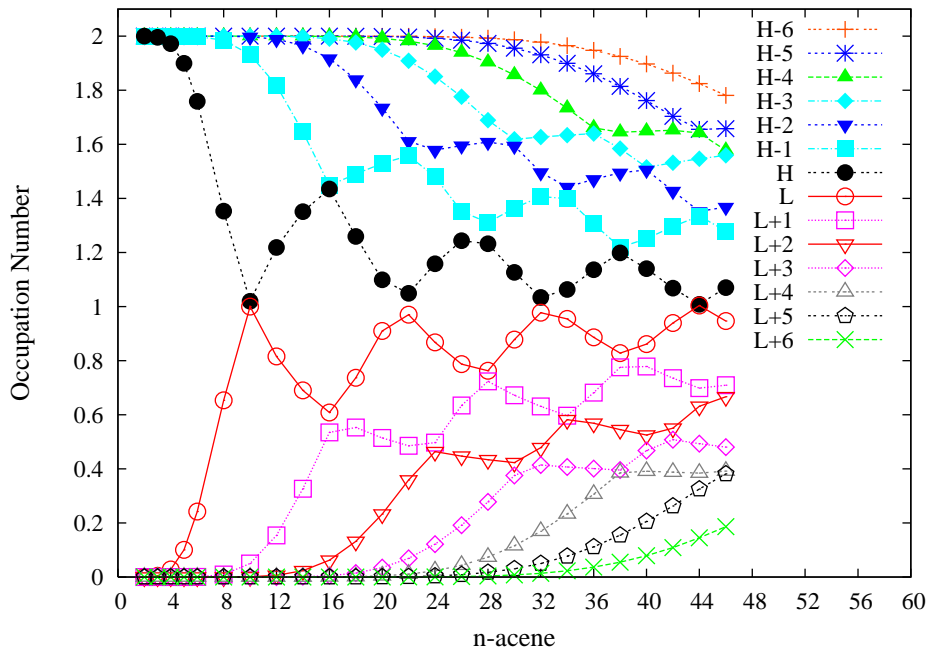}
\end{figure}

Fig.\ (\ref{acene_noon}) shows the active orbital occupation numbers for the lowest singlet states of $n$-acenes as a function of the acene length, calculated by spin-restricted TAO-LDA ($\theta$ = 7 mHartree)/6-31G* \cite{supp}. 
Here, for simplicity, HOMO, HOMO$-$1, ..., and HOMO$-$6, are denoted as $H$, $H-1$, ..., and $H-6$, respectively, while LUMO, LUMO+1, ..., and LUMO+6, are denoted as $L$, $L+1$, ..., and $L+6$, respectively. 
As can be seen, the active orbital occupation numbers exhibit oscillatory behavior in the approach to unity (singly occupied) with increasing chain length. 
The number of factionally occupied orbitals is shown to increase with the increase of chain length, which supports previous finding that large acenes should exhibit 
polyradical character \cite{aceneChan,aceneJiang}. 

To sum up, it seems plausible to believe the results obtained by TAO-LDA ($\theta$ = 7 mHartree) here, as the calculated ST gaps are in good agreement with the existing experimental and high-level {\it ab initio} 
data \cite{aceneChan,aceneHajgato,aceneHajgato2}, the calculated HOMO occupation numbers match reasonably well with the accurate NOONs, and no unphysical symmetry-breaking effects occur for 
the lowest singlet states of $n$-acenes.

\section{Definition of an Optimal $\theta$ Value}

In our study, TAO-LDA (with some fictitious temperature $\theta$) has been found to perform reasonably well for multi-reference systems, when the orbital occupation numbers $\{f_i\}$ are close to the exact NOONs $\{n_i\}$. 
In such a situation, the strong static correlation effects can be properly simulated by the entropy contribution of TAO-LDA. However, for multi-reference systems, the optimal $\theta$ for TAO-LDA has been found to be 
highly system-dependent, ranging from 5 to 50 mHartree, due to the different strengths of static correlation and the diversities of the $\{n_i\}$. On the other hand, for single-reference 
systems (in the absence of strong static correlation effects), TAO-LDA (with a $\theta$ smaller than 10 mHartree) has been shown to perform similarly to KS-LDA. 

For TAO-LDA, although it is impossible to choose a $\theta$ that is optimal for all the systems studied, it is still useful to define one to provide an explicit description of orbital occupations. Here, the optimal $\theta$ value 
is defined as the largest $\theta$ value for which the performance of the TAO-LDA (with this $\theta$) and KS-LDA is similar for single-reference systems. 
Based on our numerical investigations, an optimal value of $\theta$ = 7 mHartree, is finally chosen. TAO-LDA ($\theta$ = 7 mHartree) has been shown to consistently improve upon KS-LDA for multi-reference systems, 
while performing similarly to KS-LDA for single-reference systems.

\section{Conclusions}

We have proposed TAO-DFT, a DFT with fractional orbital occupations produced by the Fermi-Dirac distribution (in order to simulate the distribution of orbital occupation numbers for interacting electrons). TAO-DFT offers 
an explicit description of strong static correlation via the entropy contribution, a function of the fictitious temperature $\theta$ and orbital occupation numbers $\{f_i\}$ (implicit density functionals). 
Even at the simplest LDA level, the resulting TAO-LDA has been shown to perform reasonably well for multi-reference systems (due to the appropriate treatment of static correlation), when the $\{f_i\}$ (related to the $\theta$) 
are close to the exact NOONs $\{n_i\}$. As this feature is independent of the number of unpaired electrons in a system, TAO-LDA seems to be very useful for the study of large polyradical systems. 
In our study, an optimal value of $\theta$  = 7 mHartree, has been defined based on physical arguments and numerical investigations. TAO-LDA ($\theta$ = 7 mHartree), though not optimal for all the systems studied, 
has been shown to consistently improve upon KS-LDA for multi-reference systems, while performing similarly to KS-LDA for single-reference systems. Due to its computational efficiency and reasonable accuracy, 
TAO-LDA has been applied to the study of the ST gaps of acenes, which are challenging problems for conventional electronic structure methods. At the level of TAO-LDA ($\theta$ = 7 mHartree)/6-31G, 
the ST gap of polyacene has been shown to be vanishingly small (within 0.1 kcal/mol), and large acenes should exhibit singlet polyradical character in their ground states. 

As TAO-LDA is conceptually simple, computationally efficient, and easy to implement, it seems to be a promising method for the study of ground states of large single- and multi-reference systems. 
However, as with all approximate electronic structure methods, some limitations remain. The optimal $\theta$ = 7 mHartree for TAO-LDA is system-independent (not fully optimized for each system), a 
system-dependent $\theta$ (related to the distributions of NOONs) is expected to enhance the performance of TAO-LDA for a wide range of systems. For single-reference systems, the performance of TAO-LDA is similar to 
that of KS-LDA. A possible TAO-DFA is expected to perform better than TAO-LDA for single-reference systems. Although the SCEs of TAO-DFAs are expected to be less than those of KS-DFAs, 
the SIEs and NCIEs of TAO-DFAs may remain enormous in situations where these failures occur. A fully nonlocal TAO-DFT (i.e.\ nonlocal $E_{xc}[\rho]$ and $E_{\theta}[\rho]$) may be needed to resolve 
all the three qualitative errors (SIE, NCIE, and SCE). We are currently investigating along these lines, and results may be reported elsewhere.

\begin{acknowledgments}

This work was supported by National Science Council of Taiwan (Grant No. NSC98-2112-M-002-023-MY3), National Taiwan University (Grant No. 99R70304 and 10R80914-1), and NCTS of Taiwan. 
We are grateful to the Computer and Information Networking Center at NTU for the support of high-performance computing facilities. 

\end{acknowledgments}

\bibliographystyle{jcp}

\newpage

\newpage

\newpage

\newpage

\newpage

\newpage

\newpage

\newpage

\newpage

\newpage

\newpage

\newpage

\newpage

\newpage

\newpage

\newpage

\newpage

\newpage

\newpage

\newpage

\newpage

\newpage
\begin{table*}
\begin{center}
\caption{\label{table:reall} Statistical errors (in kcal/mol) of the reaction energies of 30 chemical reactions \cite{wB97X}, calculated by TAO-LDA (with various $\theta$ (in mHartree)). The $\theta=0$ case corresponds to KS-LDA.}
\begin{ruledtabular}
\begin{tabular}{lrrrrrrrrrrrr}
 \ $\theta $  &  \ \ $0$ & \ \ $1$ &  \ \ $3$ & \ \ $5$ &  \ \  $7$ &  \ \ $10$ &  \ \ $15$&  \ \ $20$&  \ \ $30$&  \ \ $40$\\
\hline
MSE&		$	-0.41	$&$	-0.72	$&$	-0.94	$&$	-1.13	$&$	-1.32	$&$	-1.59	$&$	-1.96	$&$	-2.25	$&$	-2.73	$&$	-3.04	$ \\
MAE&		$	8.51	$&$	8.27	$&$	7.75	$&$	7.36	$&$	7.09	$&$	6.95	$&$	7.53	$&$	8.92	$&$	12.28	$&$	15.21	$ \\
rms&		$	11.10	$&$	10.89	$&$	10.31	$&$	9.76	$&$	9.38	$&$	9.16	$&$	9.75	$&$	11.25	$&$	15.20	$&$	19.03	$ \\
Max($-$)& 		$	-18.31	$&$	-17.43	$&$	-15.65	$&$	-15.73	$&$	-15.92	$&$	-16.55	$&$	-18.63	$&$	-22.61	$&$	-30.65	$&$	-39.72	$ \\
Max($+$)&		$	35.68	$&$	35.59	$&$	33.88	$&$	32.18	$&$	30.50	$&$	28.01	$&$	23.95	$&$	20.08	$&$	17.09	$&$	25.45	$ \\
\end{tabular}
\end{ruledtabular}
\end{center}
\end{table*}

\newpage
\begin{table*}
\begin{center}
\caption{\label{table:EXTS} Statistical errors (in {\AA}) of EXTS \cite{EXTS}, calculated by TAO-LDA (with various $\theta$ (in mHartree)). The $\theta=0$ case corresponds to KS-LDA.}
\begin{ruledtabular}
\begin{tabular}{lrrrrrrrrrrrr}
 \ $\theta$ &  \ \ $0$ & \ \ $1$ &  \ \ $3$ & \ \ $5$ &  \ \ $7$ &  \ \ $10$ &  \ \ $15$&  \ \ $20$&   \ \ $30$&   \ \ $40$\\
\hline
MSE&		 	 	$	0.004	$&$	0.004	$&$	0.004	$&$	0.005	$&$	0.005	$&$	0.005	$&$	0.006	$&$	0.008	$&$	0.015	$&$	0.030	$\\
MAE&		 	  	$	0.013	$&$	0.013	$&$	0.013	$&$	0.013	$&$	0.013	$&$	0.013	$&$	0.013	$&$	0.014	$&$	0.021	$&$	0.036	$\\
rms&		 	 	    $	0.017	$&$	0.017	$&$	0.017	$&$	0.017	$&$	0.017	$&$	0.018	$&$	0.019	$&$	0.021	$&$	0.037	$&$	0.080	$\\
Max($-$)& 		 	 	$	-0.091	$&$	-0.091	$&$	-0.091	$&$	-0.091	$&$	-0.091	$&$	-0.092	$&$	-0.095	$&$	-0.101	$&$	-0.110	$&$	-0.110	$\\
Max($+$)&		 	 	$	0.081	$&$	0.081	$&$	0.080	$&$	0.080	$&$	0.080	$&$	0.080	$&$	0.078	$&$	0.083	$&$	0.222	$&$	0.581	$\\
\end{tabular}
\end{ruledtabular}
\end{center}
\end{table*}

\newpage
\begin{table*}
\begin{center}
\caption{\label{table:acenes3a} Singlet-triplet energy gaps (ST gaps) of $n$-acenes in the polymer limit ($n \to \infty$), obtained by nonlinear least-squares fittings of 3 different 
data sets ($20$- to $74$-acene, $30$- to $74$-acene, and $40$- to $74$-acene) of the ST gaps calculated by spin-unrestricted TAO-LDA ($\theta$ = 7 mHartree)/6-31G, using a power-law fitting function 
of the form $a + b \ n^{-c}$. Here, the coefficient of determination $R^2$ is a statistical measure of the goodness-of-fit ($R^2$ = 1, for a perfect fit).}
\begin{ruledtabular}
\begin{tabular}{cccccc}
 \ Data set&  \ \ ST gap (kcal/mol)  \ \ & \ \ $a$ (kcal/mol)  \ \ &  \ \ $b$ (kcal/mol) \ \ &  \ \ $c$  \ \ &  \ \ $R^2$ \ \ \\
\hline
$20$- to $74$-acene	&$	0.08 $&$	0.077247	\pm 0.004245 $&$	72.249	\pm 0.737 $&$	1.1199	\pm 0.0038 $&$	1.0000	$\\
$30$- to $74$-acene	&$	0.04 $&$	0.041992	\pm 0.002988 $&$	64.485	\pm 0.593 $&$	1.0806	\pm 0.0032 $&$	1.0000	$\\
$40$- to $74$-acene	&$	0.03 $&$	0.028669	\pm 0.001251 $&$	61.405	\pm 0.284 $&$	1.0645	\pm 0.0015 $&$	1.0000	$\\
\end{tabular}
\end{ruledtabular}
\end{center}
\end{table*}

\end{document}